\documentclass[journal=aamick,manuscript=article]{achemso}

\usepackage[usenames,dvipsnames,table]{xcolor}
\usepackage[version=3]{mhchem} %
\usepackage{miller}
\usepackage{subcaption}
\usepackage{multirow}
\usepackage{color}
\usepackage{textgreek}

\usepackage{amsmath,amssymb}
\graphicspath{{Figures/}}

\usepackage{pdfpages}

\usepackage[T1]{fontenc}
\usepackage[utf8]{inputenc}
\usepackage{newunicodechar}
\newunicodechar{₂}{$_2$}

\usepackage[%
  breaklinks,             %
  bookmarks = false, %
  pdfpagemode   = UseNone,%
  pdfstartview  = FitH,     %
  pdfstartpage  = 1,      %
  colorlinks    = true,   %
]{hyperref}
\hypersetup{
    linkcolor=RubineRed,          %
    citecolor=RoyalBlue,        %
    filecolor=Mulberry,      %
    urlcolor=RoyalBlue           %
}

\colorlet{mylinkcolor}{YellowOrange}
\colorlet{myurlcolor}{Aquamarine}
\colorlet{mycitecolor}{violet}



\author{Md Salman Rabbi Limon}
\affiliation{Department of Mechanical Engineering, Texas Tech University, Lubbock, Texas 79409, USA}
\author{Abrar Fahim Navid}
\affiliation{Department of Mechanical Engineering, Texas Tech University, Lubbock, Texas 79409, USA}
\author{Curtis Wesley Duffee}
\affiliation{Department of Mechanical Engineering, Texas Tech University, Lubbock, Texas 79409, USA}
\author{Zeeshan Ahmad}
\affiliation{Department of Mechanical Engineering, Texas Tech University, Lubbock, Texas 79409, USA}
\email{zeeahmad@ttu.edu}

\title{Grain Boundaries in Ceramic Solid-State Lithium Metal Batteries: A Review}


\begin{document}


\begin{abstract}
It is now widely accepted that grain boundaries play a critical role in the performance and reliability of solid-state batteries with lithium metal anodes. Understanding and controlling grain boundaries is essential for enabling safe, high-rate operation of solid-state batteries. This review explores the multifaceted influence of grain boundaries in ceramic solid electrolytes and metal anodes, including their impact on ionic and electronic transport, dendrite and void formation, connecting them to the failure mechanisms. We discuss the formation and structure of space charge layers at grain boundaries, their role in modulating local defect chemistry, and the conditions under which grain boundaries may serve as fast-ion pathways or as vulnerable sites for failure. We highlight key differences in the grain boundaries of different classes of solid electrolytes and advances in modeling, experimental characterization, and processing techniques to understand the complexity and engineer grain boundaries in solid electrolytes. Finally, we outline key open questions and opportunities for grain boundary engineering to stimulate further progress in the field.

\end{abstract}

\section{Introduction}

Lithium-ion (Li-ion) batteries have emerged as the dominant technology for electrochemical energy storage due to their high energy density, long cycle life, and reliability~\cite{Tarascon2001,Armand2008}. These batteries have revolutionized many industries, from portable electronics and electric vehicles to grid-scale energy storage. However, increasing demands for energy and power density, safety, and cyclability under harsh conditions have led to the search for newer battery chemistries. 
These demands are placed by criteria such as fast charging required for increasing electric vehicle adoption, high-rate discharging during electric flight takeoff and landing, and operation under low temperatures~\cite{Liu2019challenges,viswanathanChallengesOpportunitiesBatterypowered2022,epsteinConsiderationsReducingAviation2019}. Solid-state batteries (SSBs) offer a promising route to address these demands since solid electrolytes (SEs) have the potential to enable Li metal anodes~\cite{burtonTechnoeconomicAssessmentThin2024,janek2016solid} which have higher theoretical capacity ($>10$ times for Li compared to graphite) and lower voltage, resulting in cell energy density improvement by $\sim$50\% and specific energy improvement by $\sim$ 35\% over the current state of the art~\cite{albertusStatusChallengesEnabling2018}.
While the initial promise of SSBs was based on the discovery of ceramic SEs with conductivities rivaling those of liquid electrolytes~\cite{Kamaya2011,muruganFastLithiumIon2007}, e.g., \ce{Li10GeP2S12}, further progress has been hampered due to failure to charge/discharge them beyond the critical current density~\cite{janekChallengesSpeedingSolidstate2023a,ahmadChemomechanicsFriendFoe2022,chengIntergranularLiMetal2017,kasemchainanCriticalStrippingCurrent2019a}. 

Dendrite growth and void formation at the Li metal anode/SE have been identified as the failure mechanisms of SSBs during fast or prolonged charging and discharging, respectively. This review is motivated by the observation that both of these failure modes are significantly influenced by grain boundaries within the ceramic SE and Li metal anode~\cite{ningDendriteInitiationPropagation2023}. Notably, penetration and growth of Li dendrites through grain boundaries and the subsequent fracture have been identified as the most common failure mechanism during fast charging~\cite{porzMechanismLithiumMetal2017,renDirectObservationLithium2015}. 
Discharging at high rates leads to the formation of voids at the Li anode/SE interface, causing increased impedance. This time, grain boundaries within the Li metal play an important role in local ion transport and void formation~\cite{fuchsImagingMicrostructureLithium2024,yoonExploitingGrainBoundary2023}.  At a fundamental level, grain boundaries control two important aspects: ion transport and mechanical properties of the SE and Li anode, which affect the emergent failure event.

\begin{figure}[htbp]
    \centering
    \includegraphics[width=0.90\textwidth]{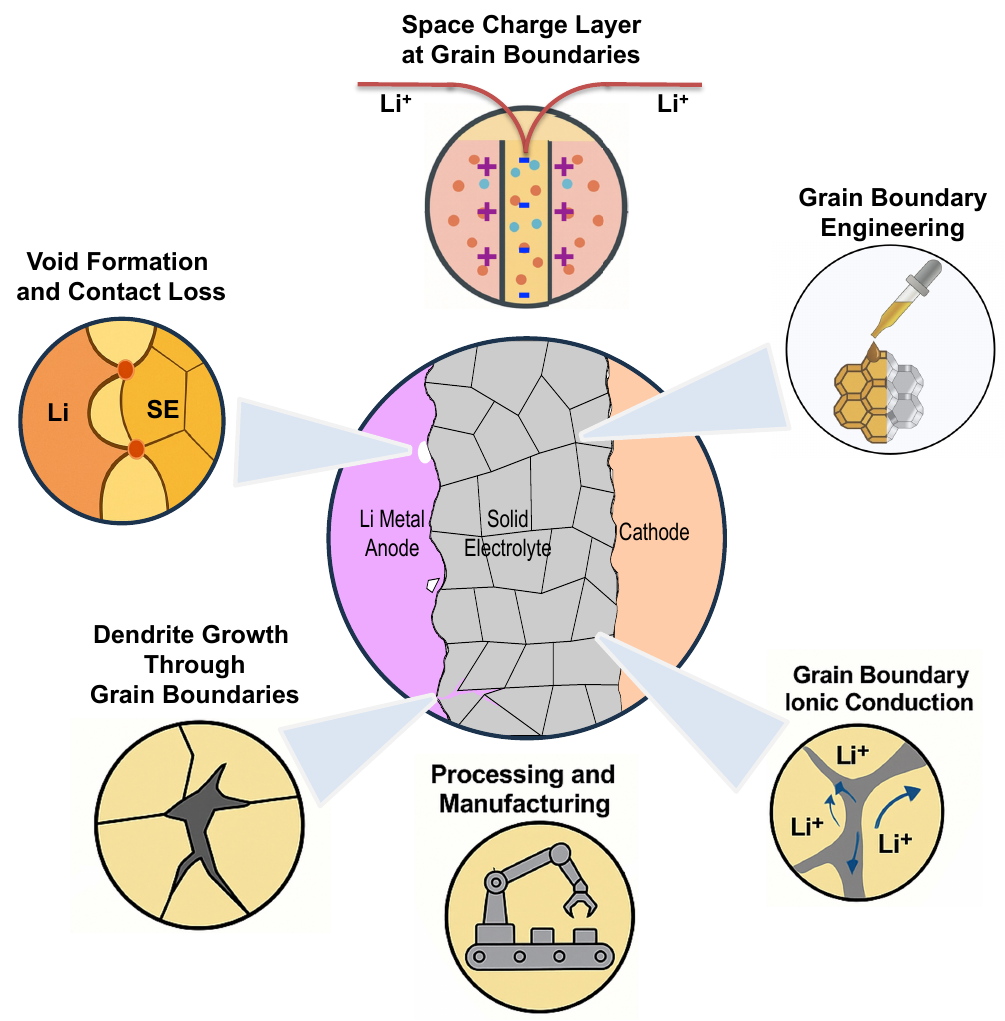} 
    \caption{Overview of key grain boundary phenomena and engineering strategies in solid electrolytes, including space charge layers, void formation, dendrite growth, ionic conduction, processing, and interface engineering, which collectively influence the performance and stability of SSBs.
    }
    \label{fig:TOC_intro}
\end{figure}
This review offers a bottom-up view on recent advances in understanding how grain boundaries within ceramic solid electrolytes and Li metal anodes influence the performance and failure mechanisms of SSBs. 
Our review complements excellent existing reviews on this subject~\cite{milanRoleGrainBoundaries2023,dawson2023going}. 
\autoref{fig:TOC_intro} illustrates the various roles and effects of grain boundaries that are reviewed in this work. One of the key phenomena associated with grain boundaries in SEs is the formation of the space charge layer, reviewed in \autoref{sec:scl}, characterized by large changes in Li ion concentration and electrical potential~\cite{maierIonicConductionSpace1995}. This layer affects the overall ionic conductivity of the SE depending on whether it is depleted or enriched with mobile \ce{Li+} defects. 
We then review the current understanding of dendrite penetration and growth through grain boundaries during metal plating in \autoref{sec:dendrite}. Void formation during fast discharging is also strongly connected with the internal microstructure of Li metal anodes and SE, as discussed in \autoref{sec:void}. 
We highlight the differences in grain boundaries of different classes of SE, namely, oxides, sulfides, halides, and antiperovskites in \autoref{sec:differences}. We conclude with a future outlook and a discussion of the key open questions regarding the role of grain boundaries in SSBs in \autoref{sec:outlook}.

\section{Space Charge Layer at Grain Boundaries}\label{sec:scl}

The space charge layer is a non-electroneutral region that forms near interfaces or grain boundaries of SEs, arising from the redistribution of point defects in response to discontinuities in chemical potential, strain, or electrostatic potential~\cite{chen2016space}. This local deviation from electroneutrality is essential to maintain electrochemical equilibrium across interfaces. The space charge layer, together with the interfacial or grain boundary core layer, forms the electrical double layer~\cite{maierIonicConductionSpace1995,ahmadUnifiedConsistentElectrical2025,swiftModelingElectricalDouble2021}. 
The accumulation or depletion of mobile ionic carriers (e.g., $\mathrm{Li^+}$) in the electrical double layer leads to modified local conductivity and potential drops~\cite{chen2016space}.

\subsection{Modeling of Grain Boundary Space Charge  Layer}

Space charge layers near grain boundaries have been modeled using a range of approaches, from classical continuum theories to fully atomistic simulations, with each framework capturing distinct aspects of interfacial behavior. While many of these frameworks focused on electrode/electrolyte interfaces, they can also be used to calculate defect concentration and potential profiles near grain boundaries. 

Continuum electrochemical models provide the foundation for describing space charge layer formation in SEs. 
Classical Poisson–Boltzmann and Mott–Schottky models yield analytical solutions for defect profiles and potential distributions across interfaces similar to liquid electrolytes~\cite{chen2016space,bardElectrochemicalMethodsFundamentals2022,maierIonicConductionSpace1995}.  
These models predict depletion widths and defect gradients that are directly applicable to polycrystals and heterostructured SEs.  
\citet{braunThermodynamicallyConsistentModel2015} advanced this framework by deriving a thermodynamically consistent model that couples electrostatics, defect chemistry, and immobile anion lattices.  
Their work revealed that solid-state space charge layers at interfaces can be orders of magnitude thicker than liquid analogs due to fixed anion frameworks and high dielectric permittivity.  
\citet{becker-steinbergerStaticsDynamicsSpaceChargeLayers2021} further extended this to describe both static and dynamic space charge evolution using coupled Poisson–Nernst–Planck equations with interfacial reaction kinetics, predicting asymmetric space charge layer formation under bias and linking interfacial charge accumulation to time‐dependent transport.  
Beyond electrostatics, Chen et al.~\cite{chenElectrochemomechanicalChargeCarrier2021} incorporated mechanical effects, demonstrating that partial molar volumes of charged defects generate stress‐dependent driving forces for defect redistribution, producing different space charge layers under bending or compressive stress.  They demonstrated that tensile stress enriches interstitial defects and enhances conduction, whereas compressive stress depletes Li$^+$ and suppresses transport.

First‐principles‐informed continuum models bridge classical electrochemical frameworks with material‐specific chemical insights obtained from first-principles simulations. An important contribution of these models is the material-specific information in terms of defect types and their properties, such as formation energy.  These results also motivated the development of models for capturing large potential drops and defect concentrations where the classical Poisson-Boltzmann framework fails.
\citet{swiftFirstPrinciplesPredictionPotentials2019} used density functional theory (DFT)‐derived defect thermodynamics and electrostatic potentials at surfaces to compute interfacial potential drops and band bending at \ce{Li2PO2N}  (LiPON)/Li and LiPON/Li$_x$CoO$_2$ interfaces.  
They incorporated these findings into a Poisson–Fermi–Dirac model that unifies the high‐defect‐density Mott–Schottky regime with the diffuse Gouy-Chapman-like regime, producing self‐consistent electrostatic and defect profiles\cite{swiftModelingElectricalDouble2021}.  
They evaluated the defect concentration and potential drops at Li/\ce{Li7La3Zr2O12}(LLZO) and Li/interlayer (LiF, Li$_2$O, Li$_2$CO$_3$) interfaces. They found that high Li‐site‐density materials (e.g., LLZO) form thin, Mott–Schottky‐like depletion layers, whereas low‐defect‐density interlayers exhibit broader, diffuse space charge layers. This contrast arises from differences in bulk defect formation energy and the saturation behavior of interfacial defects.

\begin{figure}[htbp]
    \centering
    \includegraphics[width=0.4\textwidth]{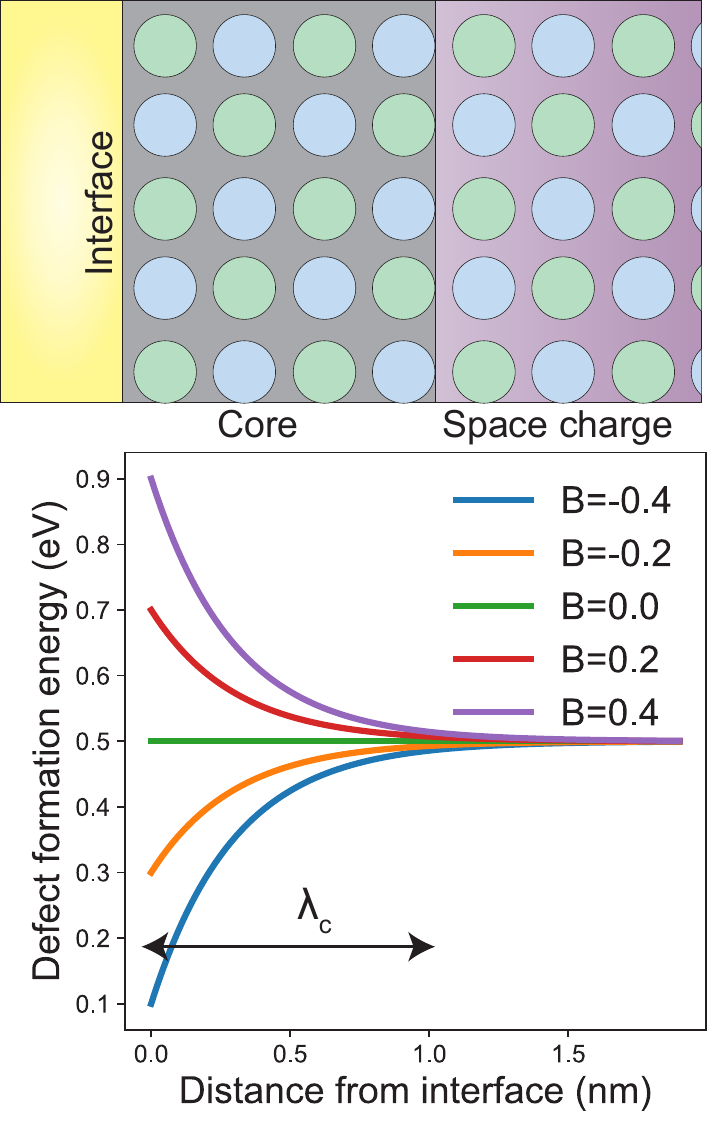} 
    \caption{
        Variation in defect formation energy with distance from an SE interface, illustrating the unified core and space charge layer treatment~\cite{ahmadUnifiedConsistentElectrical2025}. 
        A negative $B$ value reduces defect formation energy at the interface, enriching the core region with defects and promoting Li$^+$ accumulation in the adjacent space charge layer, whereas a positive $B$ leads to depletion. 
        This refined treatment highlights the role of interfacial defect formation energy in modulating ionic transport and potential drop across grain boundaries.
        Reproduced with permission from \textit{Journal of Materials Chemistry A}. Copyright ©2025, The Royal Society of Chemistry.
    }
    \label{fig:ahmad_edl}
\end{figure}

~\citet{ahmadUnifiedConsistentElectrical2025} further developed an electrical double layer model that explicitly treats both the interfacial or grain boundary core and space charge layers. 
It couples the Poisson equation with equilibrium defect thermodynamics, including spatial variation in defect formation energy near interfaces and defect–defect interactions.  
This framework captures exponential variation of defect formation energy with distance from the interface obtained from first-principles simulations~\cite{limonHeterogeneityPointDefect2024c,ahmadModulationPointDefect2024c} and reveals that core layers can dominate ionic transport when their formation energy is lower than that of the bulk.  
\autoref{fig:ahmad_edl} illustrates how interfacial defect formation energy variation modulates Li$^+$ enrichment or depletion across the core and space charge layer regions, directly linking local energetics to interfacial potential and grain boundary‐limited conductivity. 
~\citet{limonHeterogeneityPointDefect2024c} performed DFT calculations of defects near surfaces, demonstrating that local defect formation energy variations generate defect concentration gradients that could explain experimentally observed Li depletion or enrichment at grain boundaries and interfaces.

\textit{Ab initio} molecular dynamics (AIMD) simulations capture material-specific structural, dynamic, and topological effects that static models alone cannot resolve.  
\citet{dawson_atomic-scale_2018} simulated grain boundaries of \ce{Li3OCl} and found that transport was severely impeded in these regions. The activation energy for Li movement was increased by 0.11-0.27 eV at grain boundaries compared to the bulk. They further showed that grain boundary resistance dominates for grains below $\sim$100~nm, whereas bulk‐like conduction emerges above $\sim$400~nm. 
Combining DFT with machine learning interatomic potentials further enhances the capability by capturing local defect distributions, coordination environments, and electrostatic shifts at a fraction of the cost of DFT alone~\cite{dawson2023going}.  
\citet{sadowski2024grain} found through AIMD simulations that grain boundary transport in \ce{Li6PS5Br} argyrodite SE depends critically on the Br/S site exchange; higher degrees of exchange lead to reduced conductivity. This observation calls for approaches that integrate coordination chemistry and ionic topology into space charge modeling in sulfide SEs.
Further, AIMD simulations on \ce{Li_{0.33}La_{0.56}TiO}(LLTO) space charge layer near grain boundaries with the experimentally observed composition (Li excess)  confirmed that Li‐rich space charge layers maintain near‐bulk conductivity and emphasizing the importance of composition on transport~\cite{guAtomicscaleStudyClarifying2023}. A challenge that remains for standard DFT and AIMD simulations in capturing experimentally observed grain boundary properties is to equalize the electrochemical potential throughout the system in a grand canonical sense.

Overall, these multiscale modeling strategies spanning the continuum to atomistic scales demonstrate that realistic grain boundary modeling requires accurately capturing the composition (i.e., Li enrichment or depletion), electrostatic potential, defect behavior, and distortions of crystal structure at interfaces.
Such a multiscale understanding directly informs grain boundary engineering for fast-conducting SEs.

\subsection{Defect Redistribution near Grain Boundaries}

The formation of space charge layers in SEs arises from the redistribution of charged point defects near grain boundaries and heterointerfaces. Mobile species, including Li vacancies ($V_{\mathrm{Li}}^{-}$), Li interstitials ($\mathrm{Li}_{i}^{+}$), and oxygen vacancies ($V_{\mathrm{O}}^{\cdot\cdot}$), migrate under electrostatic and chemical potential gradients to compensate for fixed grain boundary charges, creating Li$^+$‐enriched or Li$^+$‐depleted regions that control interfacial transport~\cite{chen2016space}. For example, in composite SSBs, chemical‐potential differences across oxide/sulfide interfaces drive Li$^+$ accumulation or depletion, producing space charge layers whose thickness is dictated by interfacial defect energetics~\cite{chen2016space}.

\begin{figure}[htbp]
    \centering
    \includegraphics[width=1.0\textwidth]{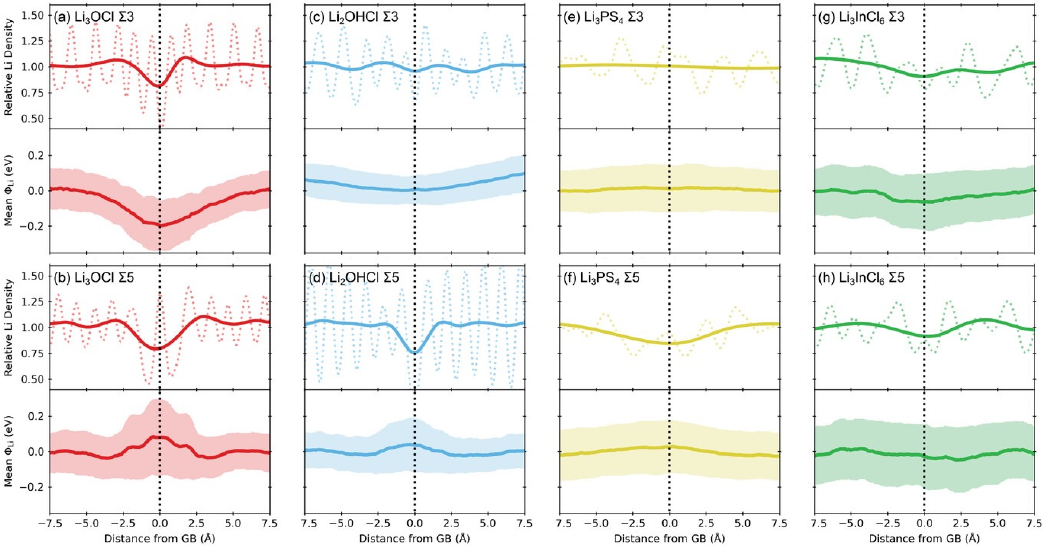}
    \caption{Normalized Li density (top) and mean local electrostatic potential $\phi_\mathrm{Li}$ (bottom) as a function of distance from grain boundaries in Li$_3$OCl, Li$_2$OHCl, Li$_3$PS$_4$, and Li$_3$InCl$_6$. Redistribution of Li ions and vacancies near grain boundary cores governs the local electrostatic landscape and space‐charge behavior. Figure reproduced from Quirk and Dawson~\cite{quirk_design_2023} under the terms of the Creative Commons Attribution License.}
    \label{fig:quirk_fig5}
    \end{figure}

The magnitude and spatial extent of defect redistribution depend on both electrostatic potential and the local defect formation energy~\cite{limonHeterogeneityPointDefect2024c,ahmadModulationPointDefect2024c,ahmadUnifiedConsistentElectrical2025}. First-principles simulations revealed that the formation energy of Li vacancies and interstitials decreases near surfaces compared to the bulk in the case of \ce{Li3OCl}, whereas it increases for crystalline LiPON~\cite{limonHeterogeneityPointDefect2024c}. Hence, the surfaces of \ce{Li3OCl} will be enriched with Li defects, while the opposite behavior is expected for LiPON. 
Similarly, ~\citet{quirk_design_2023} mapped normalized Li density and local electrostatic potential 
$\phi_\mathrm{Li}$ across grain boundaries in Li$_3$OCl, Li$_2$OHCl, Li$_3$PS$_4$, and Li$_3$InCl$_6$ (\autoref{fig:quirk_fig5}). 
Li$_3$OCl exhibits $\sim$80\% Li depletion and $\phi_\mathrm{Li}$ perturbations of 0.2\,eV ($\Sigma3\{112\}$), 
while Li$_2$OHCl shows only weak redistribution due to OH$^-$ screening. 
By contrast, Li$_3$PS$_4$ and Li$_3$InCl$_6$ display nearly flat profiles with minimal space‐charge effects. 

LLTO space charge layers have been the subject of numerous computational and experimental investigations as a prototypical material with high grain boundary resistance.  The composition of its space charge layers and reason for grain boundary resistance has been a subject of continued debate.
~\citet{sasano2021atomistic} reported that $\Sigma13$ grain boundaries accumulate oxygen vacancies, producing Li$^+$‐depleted space charge layers which increase the resistance. Wu and Guo~\cite{wuOriginLowGrain2017} observed that electron injection enhances grain boundary conductivity, consistent with compensating positive grain boundary core charges. These results were in disagreement with ~\citet{guAtomicscaleStudyClarifying2023}, who observed that LLTO grain boundary cores are negatively charged, attracting Li$^+$ and forming Li‐rich space charge layer. They demonstrated that the transport bottleneck originates from the structurally reconstructed grain boundary core, which excludes Li sites, while the Li‐rich space charge layer remains conductive. In addition, they observed that Li‐rich space charge layers extend up to 40 nm from grain‐boundary cores, far exceeding the classical $\sim$5.5 nm estimate reported by Wu and Guo~\cite{wuOriginLowGrain2017}, with excess Li occupying 3$c$ interstitial sites that are unoccupied in the bulk.
~\citet{liu2025unraveling} showed that an ultrathin Li‐deficient interface, less than 1.5 nm wide, forms at the amorphous–crystalline front during LLTO crystallization and persists after grain impingement, exhibiting an $\sim$8\% Li deficiency relative to the crystalline bulk.

Defect redistribution at sulfide space charge layers is generally weaker than oxide SEs~\cite{quirk_design_2023}. However, interfaces may still exhibit Li depletion. 
~\citet{takada2015recent} reported that during the initial charging of Li$_{1-x}$CoO$_2$|sulfide SE interfaces, Li is preferentially extracted from the electrolyte side, forming a Li-depleted space charge layer and creating an asymmetric defect landscape at the interface.

\subsection{Characterization of Grain Boundary Space Charge Layers}

Advanced characterization studies over the past decade have provided increasingly convincing evidence for the presence and impact of space charge layers at grain boundaries in SEs. These studies employed  techniques such as electrochemical impedance spectroscopy~\cite{wuOriginLowGrain2017}, scanning transmission electron microscopy (STEM)~\cite{sasano2021atomistic,guAtomicscaleStudyClarifying2023}, atomic force microscopy (AFM), electron energy loss spectroscopy (EELS)~\cite{liu2025unraveling,guAtomicscaleStudyClarifying2023}, Kelvin probe force microscopy (KPFM)~\cite{zhuUnderstandingEvolutionLithium2023}, and nuclear magnetic resonance (NMR)~\cite{cheng2020revealing}. These techniques span atomic to macroscale and together illuminate how space charge layers originate, evolve, and influence ionic conduction.

Early experimental evidence of space charge layers came from electrochemical impedance spectroscopy measurements on SEs with varying grain sizes. 
~\citet{wuOriginLowGrain2017} investigated LLTO with different grain sizes and observed that 1) the grain boundary conductivity was four orders of magnitude lower than the bulk conductivity; 2) the grain boundary conductivity decreased with decreasing grain size, consistent with the classical brick layer conduction model~\cite{maierIonicConductionSpace1995,fleig1999impedance}. The decreased conductivity was attributed to Li depletion by over five orders of magnitude in the space charge layer. This provided a clear experimental correlation between microstructure and space charge layer-induced transport behavior at the grain boundaries, even in the absence of a second intergranular phase.

Subsequent studies introduced spatially resolved electrochemical mapping to understand grain boundary conductivity.  
~\citet{sasano2021atomistic} combined atomic-resolution STEM-imaging and AFM to investigate how specific grain boundary structures in LLTO affect Li-ion conductivity. Using KPFM and electrochemical strain microscopy (ESM), they mapped surface potential and ionic transport across Σ5 and Σ13 grain boundaries. At the Σ13 boundary, they observed positively charged oxygen vacancies and adjacent Li-ion depletion regions, which they identified as the origin of suppressed conductivity, consistent with the formation of a space–charge–like region.

\begin{figure}[htbp]
    \centering
    \includegraphics[width=0.75\textwidth]{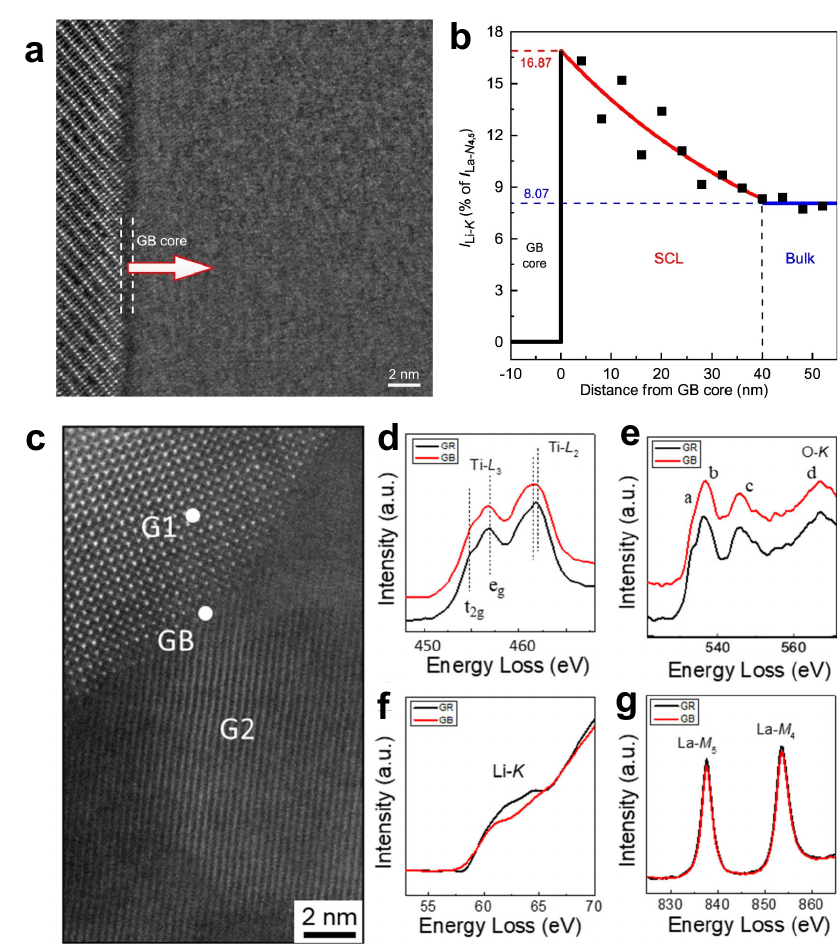}
    \caption{(a) HAADF-STEM image and 
    (b) normalized EELS line scan (Li-K relative to La-N$_{4,5}$) across an LLTO grain boundary reveal Li enrichment in the vicinity of the grain boundary. 
    The Li content peaks near the grain boundary core (exceeding 16\%) and decays exponentially over a wide spatial extent ($\sim$40\,nm). 
    These observations indicate a negatively charged grain boundary core that attracts Li$^+$, forming a Li-rich space charge layer that retains comparable ionic conductivity to the bulk. 
    This finding challenges the prevailing assumption of Li-deficient, resistive space charge layers and shifts the transport bottleneck to the Li-depleted grain boundary core itself.
    (c–g) Evidence for Li-deficient, positively charged grain boundary cores from Liu et al.~\cite{liu2025unraveling}:
    (c) Atomic-resolution HAADF-STEM image of an LLTO grain boundary between two grains (G1 and G2), identifying the grain boundary core. 
    (d–g) EELS spectra normalized to Ti-L$_{2,3}$ edges show: 
    (d) reduced $t_{2g}/e_g$ splitting, 
    (e) diminished O-K prepeak intensity, 
    (f) decreased Li-K edge intensity indicating Li deficiency, and 
    (g) unaltered La-M edge intensity. 
    These signatures support a structurally reconstructed grain boundary core that induces a Li$^+$-depletion-type layer confined to a narrow region ($<1.5$\,nm), contributing to grain boundary blocking in LLTO. (f–g) Evidence for Li-rich, negatively charged space charge layers from Gu et al.~\cite{guAtomicscaleStudyClarifying2023}
    Panels (a–b) adapted under Creative Commons License from Gu et al.~\cite{guAtomicscaleStudyClarifying2023}
    Panel (c-g) adapted with permission from Liu et al.~\cite{liu2025unraveling}, Copyright \copyright\ 2025 American Chemical Society. 
    }
    \label{fig:liu2025_gb}
\end{figure}

While earlier works attributed low grain boundary conductivity to Li depletion in the space charge layer, \citet{guAtomicscaleStudyClarifying2023} provided an alternative explanation through a combination of STEM imaging (\autoref{fig:liu2025_gb}a) and spatially-resolved EELS with Li-K/La-N$_{4,5}$ normalization. Contrary to long-standing assumptions, they found that the grain boundary core is negatively charged, resulting in Li$^+$ enrichment, not depletion. The normalized Li-K intensity more than doubled near the grain boundary, then decayed exponentially over a $\sim$40 nm region, consistent with space charge layer theory but revealing a Li-rich rather than Li-deficient character as shown in \autoref{fig:liu2025_gb}b. High-resolution imaging confirmed that this excess Li occupies interstitial 3$c$ sites. The low grain boundary conductivity was attributed to Li depletion in the core region. This finding aligns with Ahmad’s electrochemical model, which emphasizes the importance of the core layer for grain boundary conductivity~\cite{ahmadUnifiedConsistentElectrical2025}.

In situ S/TEM annealing of amorphous LLTO enabled real-time tracking of grain-boundary formation during crystallization\cite{liu2025unraveling}. 
The STEM images and EELS revealed that a Li-deficient ultrathin layer ($<1.5$ nm) spontaneously forms at the crystallization front and persists as two grains merge. \autoref{fig:liu2025_gb}c shows an atomic-resolution HAADF-STEM image of an LLTO grain boundary between two grains (G1 and G2), pinpointing the grain boundary core. In \autoref{fig:liu2025_gb}d, the EELS spectrum reveals a slight Ti-L$_{2,3}$ energy shift with reduced $t_{2g}/e_g$ splitting, while \autoref{fig:liu2025_gb}e displays a diminished O-K prepeak intensity. \autoref{fig:liu2025_gb}f shows a weakened Li-K edge at $\sim 60$~eV, indicating Li deficiency at the grain boundary core, and \autoref{fig:liu2025_gb}g confirms that the La-M edge remains unaltered. Together, these signatures indicate a structurally reconstructed, Li-deficient grain boundary core that induces a Li-depletion-type space charge layer confined to a narrow region ($< 1.5$~nm). This study, however, did not probe the Li distribution in the space charge layer.

Complementary insights into space charge layer-induced transport barriers were provided by \citet{cheng2020revealing}, who used two-dimensional exchange NMR to quantify Li$^+$ exchange across SE-cathode interface in solid-state batteries. In \ce{Li_{1.5}Al_{0.5}Ge_{1.5}(PO3)4}/Li$_x$V$_2$O$_5$ bilayers, they observed that introducing a space charge layer raised the activation energy for interfacial Li$^+$ exchange from 0.315 to 0.515 eV and suppressed the exchange current density by over an order of magnitude. These effects, supported by electrostatic modeling, were attributed to charge separation across the interface, demonstrating that space charge layers at buried interfaces can critically impede ionic transport even in the absence of chemical decomposition or structural mismatch.

Operando KPFM measurements on LLZO during Li plating showed a marked drop in the Galvani potential at grain boundaries, indicative of the space charge layer~\cite{zhuUnderstandingEvolutionLithium2023}. However, this drop was only observed for grain boundaries located near the Li counter electrode (plating side) caused due to the accumulation of electrons on that side. This study highlighted how space charge layers may be modified under applied external voltage conditions.

Collectively, these complementary experimental approaches from surface-sensitive imaging to macroscopic impedance spectroscopy, spectroscopic core-level mapping, and atomic-resolution in situ imaging form a comprehensive suite of tools to probe different features of space charge layers. They confirm that space charge layers are structurally and chemically diverse, and that their influence on ionic transport is tightly coupled to the local grain boundary chemistry and microstructural evolution.

\subsection{Strategies to Boost Space Charge Conductivity}
Space charge layers can either suppress or enhance ionic transport, depending on whether they deplete or enrich mobile charge carriers. 
They can be deliberately engineered into fast‐ion channels through “doping‐like” strategies that modulate interfacial charge and defect chemistry. The impact of space charge layers is also state-dependent. Interfacial fields can either oppose or assist Li$^+$ motion depending on the cathode state of charge, with Li-deficient space charge layers at high state of charge increasing resistance~\cite{swiftFirstPrinciplesPredictionPotentials2019,cheng2020revealing}.

Direct electronic or chemical doping suppresses Li$^+$‐depleted space charge layers.
In LLTO with highly resistive grain boundaries, Nb doping enhances grain boundary conductivity due to increased Schottky barrier height~\cite{wuOriginLowGrain2017}. 
Interfacial and composite approaches extend this principle: Chen and Guo~\cite{chen2016space} demonstrated that interlayers and compositional gradients flatten electrochemical potential profiles.
\citet{zhu_reduced_2020} introduced amorphous grain boundary domains in LLZO thin films by doping with Ga and extra Li that can diminish space charge effects and noticeably increase conductivity. 
Surface and processing strategies can further optimize ionic transport in space charge layers. 
Ultrathin oxide coatings (LiNbO$_3$, Li$_2$SiO$_3$, Li$_4$Ti$_5$O$_{12}$) can be used to suppress Li depletion at cathode-SE interfaces~\cite{takada2015recent,haruyamaSpaceChargeLayer2014}.

Several efforts have aimed to enhance SE conductivity by introducing new phases that form heterointerfaces with favorable space charge layers, which can potentially improve SE conductivity by orders of magnitude. In these cases, the sluggish ion transport through grain boundaries is replaced by faster transport through heterointerfaces. 
One of the earliest studies demonstrated that doping the poorly conducting LiI with \ce{Al2O3} enhanced its ionic conductivity by more than an order of magnitude.\cite{liangConductionCharacteristicsLithium1973}. 
Conducitivity enhancements have also been observed in mixed SEs such as \ce{CaF2}-\ce{BaF2} heterolayers~\cite{Sata2000mesoscopic}, LiCl/FeOCl~\cite{ohtaFastIonicConduction2025}, and LiF/\ce{Li2TiF6}~\cite{qianUnlockingLithiumIon2024}.
Wang et al.~\cite{wang1+122025b} demonstrated that physically mixing Li$_2$ZrCl$_6$ with Li$_3$YCl$_6$ generates interfacial space charge layers enriched with defects, producing a 46\% conductivity increase, indicating that engineered space charge layers can function as fast ion channels in composites.

\begin{figure}[htbp]
    \centering
    \includegraphics[width=0.9\textwidth]{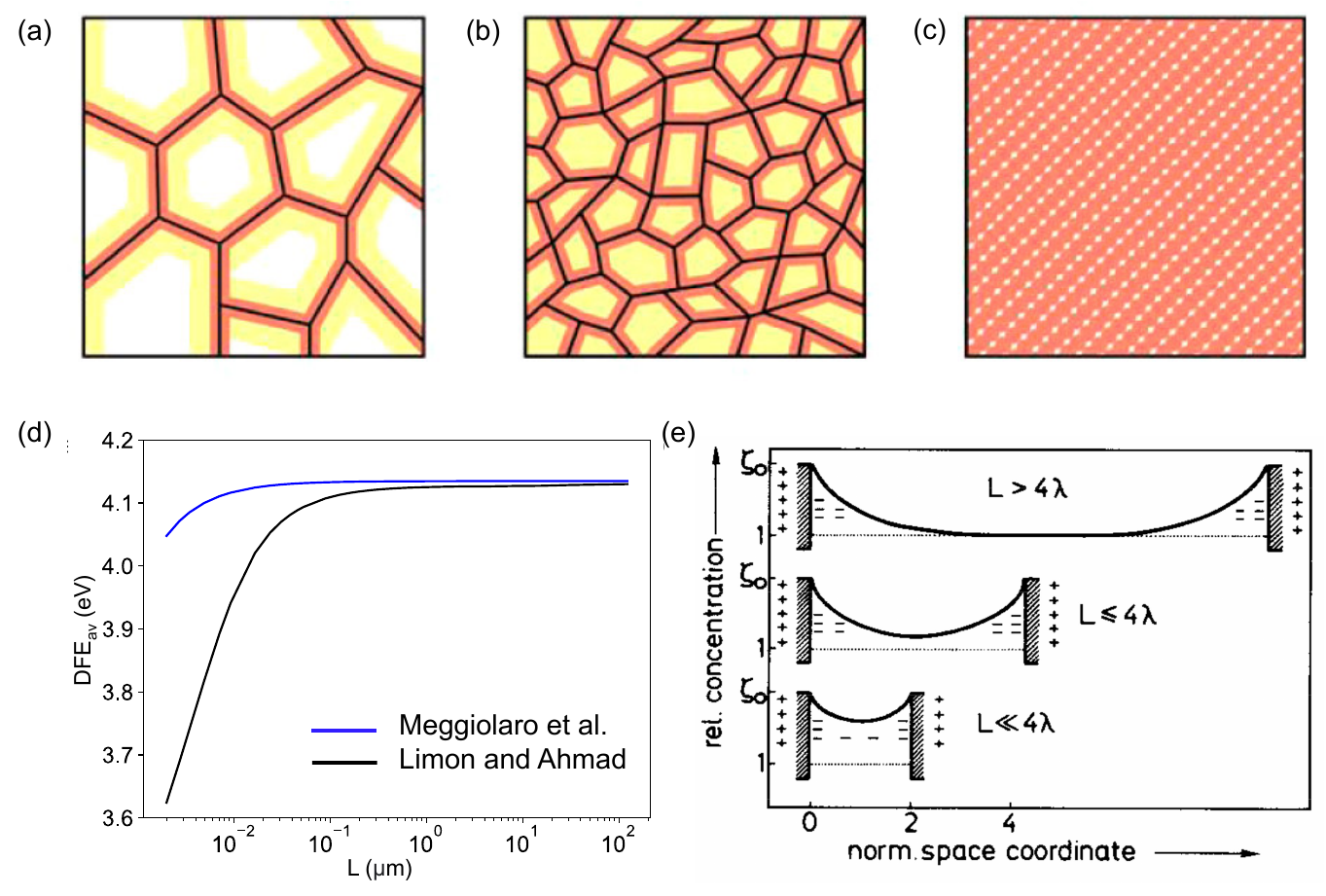}
    \caption{(a–c) Top panels illustrate how space charge layers evolve with decreasing grain size, originally conceptualized by Maier~\cite{maier2014pushing}. In moderately thin systems (a)space charge layers are confined to near-surface regions, while in very thin (b) and extremely thin (c) grains, the surface potential influences the entire volume, suppressing bulk-like behavior.  
    (d) Bottom left panel, adapted from Limon and Ahmad~\cite{limonHeterogeneityPointDefect2024c}, compares the average Li$^+$ vacancy defect formation energy in Li$_3$OCl as a function of grain size using two models: their own exponential spatial defect formation energy profile (black line) and a stepwise model from Meggiolaro et al.~\cite{meggiolaroFormationSurfaceDefects2019} (blue line), which treats only the surface as distinct from the bulk. The exponential model more accurately captures the dominance of surface defect formation energy in small grains, reflecting gradual spatial heterogeneity in defect energetics. This shows that when the grain size is comparable to the Debye length, the system cannot recover bulk-like energetics and remains governed by the lower surface defect formation energy.  
    (e) Bottom right panel, redrawn from Maier~\cite{maierDefectChemistryIonic1987}, shows schematic defect concentration profiles in thin ionic conductor films. Three thickness regimes are illustrated: moderately thin ($L > 4\lambda$), where bulk concentration is recovered in the interior; thin ($L \sim 4\lambda$), where the interior begins to polarize; and extremely thin ($L \ll 4\lambda$), where the defect redistribution spans the full film thickness, eliminating any bulk-like region. These profiles illustrate how the spatial extent of space charge regions relative to the Debye length $\lambda$ controls defect distributions. This framework supports the interpretation by Limon and Ahmad that surface-dominated energetics emerge naturally in nanoscale grains, where $L$ is on the order of or smaller than $4\lambda$. Panels (a–c) adapted with permission from Maier~\cite{maier2014pushing}, Copyright \copyright\ 2014 American Chemical Society. Panel (d) adapted with permission from Limon and Ahmad~\cite{limonHeterogeneityPointDefect2024c}, Copyright \copyright\ 2024 American Chemical Society. Panel (e) adapted with permission from Maier~\cite{maierDefectChemistryIonic1987}, Copyright \copyright\ 1987 Elsevier.
    }
    \label{fig:grain_size}
\end{figure}
Theoretical models have clarified how space charge layer characteristics vary with grain size and defect energetics. As shown in \autoref{fig:grain_size}a-c, when grain size $L$ becomes comparable to or smaller than four times the Debye length ($L \lesssim 4\lambda$), space charge effects extend across the entire grain, eliminating any bulk-like interior and resulting in a fully perturbed defect distribution~\cite{maier2014pushing}. \autoref{fig:grain_size}(d) further illustrates how average Li$^+$ vacancy formation energy in Li$_3$OCl varies with grain size. The exponential model developed by Limon and Ahmad~\cite{limonHeterogeneityPointDefect2024c} captures the smooth spatial variation of defect formation energies, unlike stepwise models~\cite{meggiolaroFormationSurfaceDefects2019} that treat only surface sites as distinct. The result is a clear trend: in small grains ($L \lesssim 4\lambda$), surface energetics dominate the average defect formation energy (and thereby the defect concentration), suppressing bulk behavior. \autoref{fig:grain_size}(e) reinforces this by showing how thin-film concentration profiles evolve with $L/\lambda$, ranging from localized space charge layers at large $L$ to fully redistributed profiles at small $L$~\cite{maierDefectChemistryIonic1987}. Tuning the interfacial defect formation energy drop and core thickness can enhance grain boundary conductivity by over 50\%, guiding integrated approaches that combine doping, interfacial engineering, and process‐controlled Li supply~\cite{ahmadUnifiedConsistentElectrical2025}. 
Together, these strategies convert space charge layers from passive barriers into functional fast‐ion channels, enabling superior ionic transport in SSBs.

\clearpage
\section{Dendrite Growth Through Grain Boundaries}\label{sec:dendrite}

Ceramics were initially identified as promising SE candidates for SSBs due to their robust mechanical properties, which can prevent dendrite penetration.  LLZO has a high shear modulus ($\sim$60 GPa) and Young’s modulus (>50 GPa), which theoretically should be enough strength to resist Li metal intrusion~\cite{yuElasticPropertiesSolid2016,dengElasticPropertiesAlkali2015}. The Monroe-Newman model applied linear stability analysis for dendrite initiation during Li metal electrodeposition and proposed that SEs with high shear modulus can suppress dendrite growth~\cite{monroeImpactElasticDeformation2005,monroeEffectInterfacialDeformation2004a}. While experiments supported this result for polymer SEs~\cite{balsara2012-modadh}, the model was found to be too simplistic for ceramic SEs due to dendrite growth through grain boundaries, surface flaws, and Li molar volumes~\cite{renDirectObservationLithium2015,porzMechanismLithiumMetal2017,ahmadStabilityElectrodepositionSolidSolid2017}.  The model was later refined to incorporate the molar volume of Li ions within ceramic SEs~\cite{ahmadStabilityElectrodepositionSolidSolid2017}. This analysis predicted a density-driven stable electrodeposition regime, suggesting that softer SEs are more effective at preventing dendrite initiation. However, it did not account for surface imperfections such as defects and cracks, where dendrites eventually grow beyond a critical current density in single-crystalline ceramics~\cite{swamyElectrochemicalChargeTransfer2015,ningDendriteInitiationPropagation2023}. Later models incorporated other ceramic properties such as fracture toughness, interfacial defect size, permittivity, and grain size to predict the critical current density for plating~\cite{liuStabilityCriterionElectrodeposition2024,liDendriteNucleationLithiumconductive2019,rajCurrentLimitDiagrams2017}.

The microstructure and surface properties of SEs have been identified as more important descriptors of dendrite growth through ceramic SEs compared to bulk properties such as shear modulus. Overall, two mechanisms have been identified for dendrite formation and growth in multiple studies, both associated with imperfections such as grain boundaries. The first mechanism involves dendrite growth through grain boundaries and cracks within the ceramic, leading to fracture and is related to the local mechanical properties. The second mechanism involves electronic leakage in the ceramic through the SE and is related to the electronic conductivity at grain boundaries. The dominant mechanism has been a subject of continued debate~\cite{hanHighElectronicConductivity2019,mcconohyMechanicalRegulationLithium2023} and the two mechanisms may be linked. Below, we discuss the two mechanisms separately.

\subsection{Mechanical Properties}
\begin{figure}[htbp]
    \centering
    \includegraphics[width=0.75\textwidth]{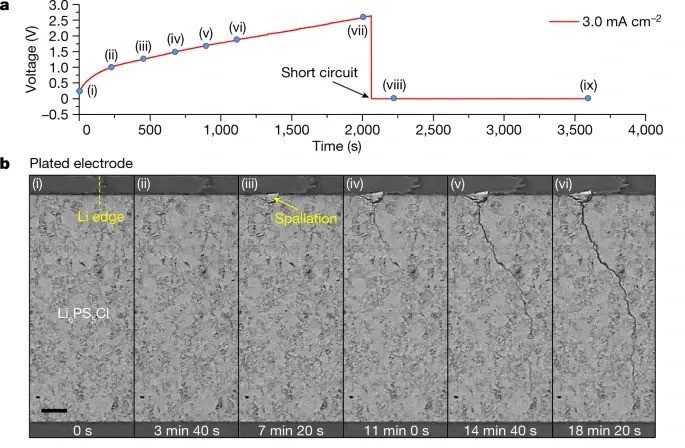}
    \caption{a) Graph showing the voltage versus time during cycling of the cell. With time, the voltage rises until the cell fails and the voltage abruptly goes to zero. b) Imaging of the crack and dendrite formation as cycling continues. As the dendrite grows, the pressure causes more cracking until the failure of the cell.~\cite{ningDendriteInitiationPropagation2023}  Reproduced with permission from Ref. \cite{ningDendriteInitiationPropagation2023}. Copyright \textcopyright 2023 Springer Nature.}
    \label{fig:ning}
\end{figure}

Many studies have reported that Li dendrites preferentially nucleate and propagate along grain boundaries of LLZO~\cite{chengIntergranularLiMetal2017,renDirectObservationLithium2015}. 
~\citet{chengIntergranularLiMetal2017} used scanning electron microscopy to image the cross-section of a pellet after failure during cycling. They found a web-like structure of the Li dendrites propagating transgranularly across the SE.
The short-circuiting time of the ceramic increased with pellet density. However, other studies did not find a clear correlation between pellet density and the occurrence of a short circuit~\cite{ishiguroStabilityNbDopedCubic2013,sudoInterfaceBehaviorGarnettype2014,tsaiLi7La3Zr2O12InterfaceModification2016}. 
 While some reports suggest that certain grain boundary chemistries can enhance \ce{Li} mobility, the net effect of inhomogeneities in ionic and electronic properties is a non-uniform deposition environment that favors dendrite growth along grain boundary pathways.~\cite{porzMechanismLithiumMetal2017}
\citet{kazyakLiPenetrationCeramic2020} conducted a comprehensive analysis of Li penetration in LLZO using optical and electron microscopy and found multiple Li growth morphologies in the same cell, some of which involved preferential growth along grain boundaries.  Another study argued that dendrite penetration probability follows a Weibull distribution with cracks forming along both the grain boundaries and the interior~\cite{mcconohyMechanicalRegulationLithium2023}. 
Operando X-ray computed tomography on \ce{Li6PS5Cl} SE revealed that dendrite initiation is based on existing microcracks and pores within the interphase\cite{ningDendriteInitiationPropagation2023}. As more Li builds up in the area, the pressure increases beyond the grain boundary fracture strength, leading to cracks where dendrites continue to propagate. Eventually, the propagating Li-filled crack forms a direct electronic connection between the two electrodes, leading to a short circuit as shown in \autoref{fig:ning}. Amorphous SEs did not exhibit this failure mode even at high current densities, indicating the prominent role of grain boundaries in dendrite formation~\cite{porzMechanismLithiumMetal2017,sakuda_sulfide_2013}.

\begin{figure}[htbp]
    \centering
    \includegraphics[width=\textwidth]{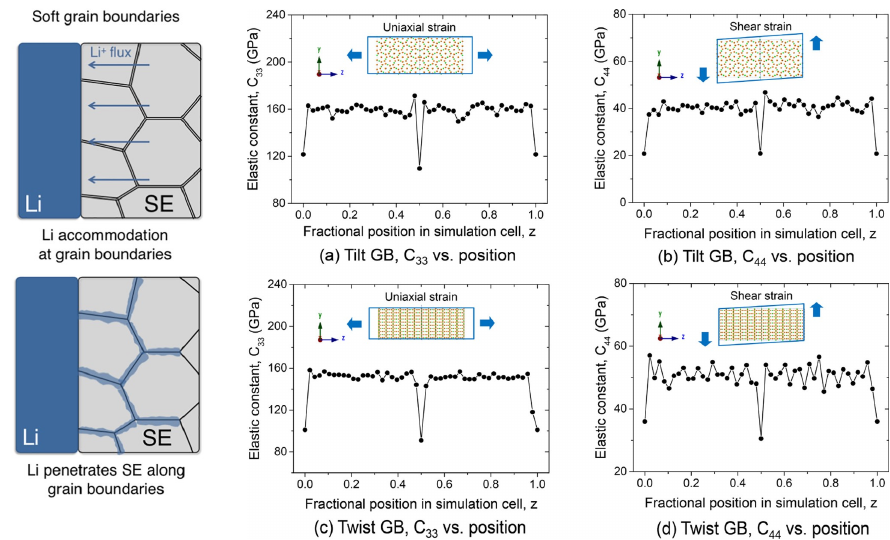}
    \caption{(Left) Schematic illustration of Li plating at soft grain boundaries in SEs, showing Li$^+$ flux during the initial stage and subsequent penetration of electrodeposited Li along grain boundaries. (Right) Spatial variation of the elastic constants $C_{33}$ and $C_{44}$ at 300~K across the $\Sigma5$ grain boundary cells: (a,b) symmetric tilt grain boundary and (c,d) twist grain boundary, evaluated as a function of fractional position normal to the grain boundary plane. Reproduced with permission from Ref.~\cite{yuGrainBoundarySoftening2018}. 
    Copyright \textcopyright 2018, American Chemical Society.}
    \label{fig:yu_siegel_soft}
\end{figure}
Considering their importance in dendrite suppression, numerous experimental and computational studies have investigated the mechanical properties of SEs and their relation with grain size and grain boundaries.
Atomistic studies on grain boundaries showed that the shear modulus of the grain boundary can be up to 50\% lower than the surrounding bulk regions, referred to as grain boundary softening, as shown in \autoref{fig:yu_siegel_soft}.~\cite{yuGrainBoundarySoftening2018} The grain boundary network represents a mechanically weaker phase within the SE that is more compliant and prone to deformation or fracture under stress. Such inhomogeneities in elastic properties provide a mechanistic pathway for soft Li metal to infiltrate stiff SEs. Using first-principles calculations, \citet{xie_effects_2024} built a database of mechanical properties of 590 surface and grain boundary models across a range of SE chemistries, including oxides, sulfides, thiophosphates, and halides. Their results showed that grain boundaries are significantly easier to separate mechanically than the bulk, suggesting a greater tendency for crack propagation along these interfaces.

The fracture toughness of the SE critically affects the dendrite growth direction. 
~\citet{Zhang2025} found that while the dendrites can initiate anywhere, they are preferentially deflected towards grain boundaries. This is caused by the lower fracture toughness of the grain boundary~\cite{wolfenstineCriticalGrainSize1999}, breaking first when the pressure is built up with the Li metal deposition\cite{chengIntergranularLiMetal2017}. 
\citet{jiao2024grain} employed an electrochemomechanical model with Butler-Volmer kinetics to explore failure mechanisms in polycrystalline SEs, with particular emphasis on the role of grain boundaries and grain size. 
Their results showed that stress accumulation at the Li/SE interface causes grain boundary opening and sliding, which facilitates the outward propagation of mechanical damage along grain boundary networks. 

\begin{figure}[htbp]
    \centering
    \includegraphics[width=0.7\textwidth]{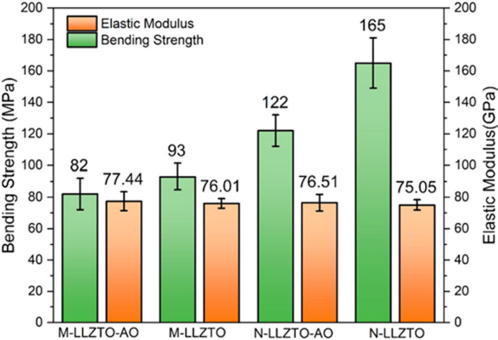}
    \caption{Bending strength and elastic modulus of LLZTO SEs 
    with different grain sizes and grain boundary chemistries. 
    M-LLZTO and N-LLZTO correspond to micro- and nano-grained samples, respectively, 
    while the “-AO’’ designation indicates the addition of Al$_2$O$_3$ during sintering. 
    Nano-grained LLZTO without additive exhibited the highest bending strength 
    ($\sim$165~MPa), while Al$_2$O$_3$-containing samples showed slightly lower 
    strengths (122~MPa for N-LLZTO-AO and 82~MPa for M-LLZTO-AO), 
    attributed to LiAlO$_2$ formation at grain boundaries. 
    The elastic modulus remained nearly constant (75--77~GPa) for all compositions. 
    Reproduced with permission from Ref.~\cite{huangEnhancedPerformanceLi2020}. 
    Copyright \textcopyright 2020, American Chemical Society.}
    \label{fig:huang_mech}
\end{figure}

~\citet{huangEnhancedPerformanceLi2020} systematically evaluated the mechanical properties of LLZTO SEs with varying grain sizes and grain boundary chemistries using three-point bending tests. As shown in ~\autoref{fig:huang_mech}, the bending strength exhibited a strong dependence on grain size: the nano-grained LLZTO without Al$_2$O$_3$ additive reached the highest value of $\sim$165~MPa, while the nano-grained sample with Al$_2$O$_3$ additive reached 122~MPa. In contrast, micro-grained samples displayed lower bending strengths of 93~MPa (without additive) and 82~MPa (with additive). This trend reflects the classical fine-grain strengthening effect in polycrystalline ceramics, where smaller grains promote crack deflection and longer fracture paths. In comparison, the elastic modulus showed minimal variation across all four samples (75-77~GPa), highlighting that grain boundary modification mainly influences fracture strength rather than stiffness. The nano-grained LLZTO with LiAlO$_2$-modified grain boundaries exhibited the highest critical current densities due to the combination of high fracture strength and lower electronic conductivity of grain boundaries.

\begin{figure}[htbp]
    \centering
    \includegraphics[width=1.0\textwidth]{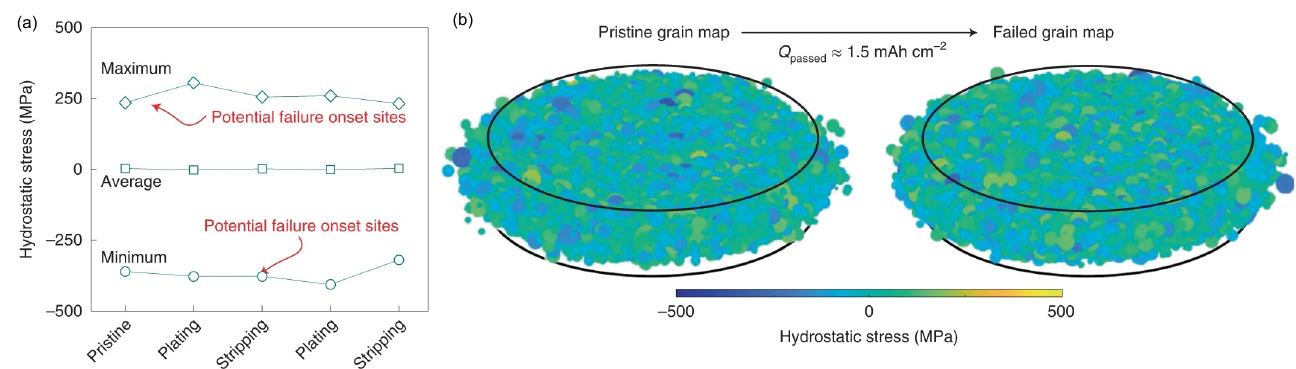}
    \caption{(a) Hydrostatic stress evolution in a Li$\vert$LLZO$\vert$Li symmetric cell 
    during plating and stripping, showing the average, maximum, and minimum stresses 
    across $\sim$30{,}000 grains. Red arrows indicate localized high- and low-stress 
    regions that act as potential failure initiation sites. 
    (b) Three-dimensional reconstructed grain maps of the LLZO pellet in the pristine 
    and failed states, color-coded by hydrostatic stress. 
    Hot- and cold-spot grains coincide with regions susceptible to intergranular fracture 
    and Li filament penetration. 
    Reproduced with permission from Ref.~\cite{dixitPolymorphismGarnetSolid2022}. 
    Copyright \textcopyright 2022, Springer Nature.}
    \label{fig:polymorphism}
\end{figure}

In addition to mechanical heterogeneity caused by grain boundaries, grain-to-grain stress variations have also been observed in SEs during cycling. \citet{dixitPolymorphismGarnetSolid2022} employed far-field high-energy diffraction microscopy combined with X-ray tomography to track the grain-level chemo-mechanical response of 30,000 grains within polycrystalline LLZO during Li plating and stripping~\cite{dixitPolymorphismGarnetSolid2022}. As shown in ~\autoref{fig:polymorphism}a and b, while the average hydrostatic stress across the pellet remained relatively constant, the maximum and minimum grain-level stresses fluctuated, revealing localized regions 
of tensile and compressive stress that served as potential failure initiation sites. These observations demonstrate that mechanical failure in garnet SEs is highly stochastic and governed by local microstructural heterogeneity due to grains and structural polymorphs.

\subsection{Electronic Conductivity}
Ideally, SEs should be ionic conductors and electronic insulators to force electrons to move only via the external circuit. In practice, however, many popular SEs exhibit a small but non-zero electronic conductivity that can undermine cell safety. For example, LLZO is a fast Li-ion conductor but has an intrinsic electron conductivity on the order of 10$^{-7}$–10$^{-8}$ S cm$^{-1}$.~\cite{Zhang2024} Such leakage may seem minimal, but under prolonged cycling, it can enable internal redox side-reactions. Electrons that stray into the SE can meet Li ions, effectively plating metallic Li within the ceramic. This mechanism has been directly linked to dendrite initiation inside the SE as opposed to the interface. \cite{hanHighElectronicConductivity2019,tsaiLi7La3Zr2O12InterfaceModification2016,songRevealingShortCircuitingMechanism2019} 
\citet{hanHighElectronicConductivity2019} demonstrated that LLZO’s modest electronic conductivity is largely responsible for internal dendrite formation, as it allows \ce{Li+} to be prematurely reduced to \ce{Li^0} within the SE. In comparison, LiPON has a lower electronic conductivity and does not exhibit dendrite growth.

\begin{figure}[htbp]
    \includegraphics[width=1\textwidth]{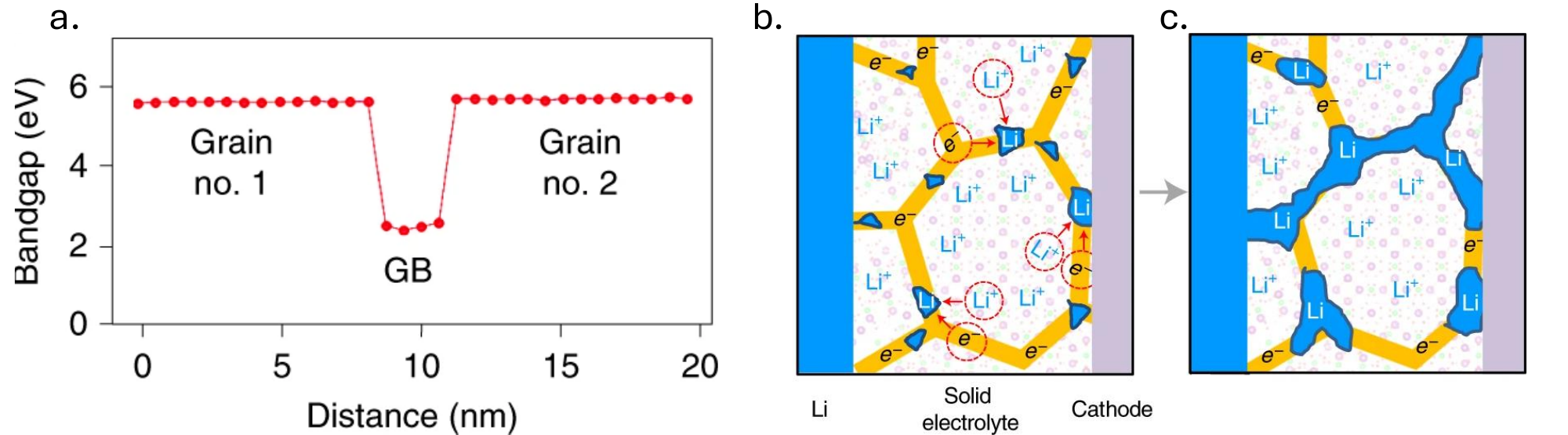}
    \caption{(a) Bandgap values extracted from the EELS line-scan across the grain boundary for LLZO. (b) A supplementary pathway for filament growth and intrusion in SSBs utilizing polycrystalline electrolytes~\cite{liuLocalElectronicStructure2021}.  Reproduced with permission from Springer Nature} %
    \label{fig:nav-electronic-conductivity}
\end{figure}
Many follow-up studies revealed that inhomogeneity in electronic properties plays an important role in dendrite nucleation and growth regions.
\citet{liuLocalElectronicStructure2021} reported that grain boundary regions in LLZO often have a significantly smaller band gap as shown in \autoref{fig:nav-electronic-conductivity}a  ($\sim$1–3 eV compared to $\sim$6 eV in the bulk) and thus, act as local electron-conduction pathways. These conductive pathways let electrons bypass the intended circuit and deposit Li metal inside the SE as shown in \autoref{fig:nav-electronic-conductivity}b and c.~\cite{liuLocalElectronicStructure2021} This increased electronic conductivity within the grain boundary is one of the driving forces for Li dendrites to propagate through these pathways instead of the grains. \citet{songRevealingShortCircuitingMechanism2019} also observed Li plating within the garnet ceramic on the grain boundaries due to higher local electronic conductivity in this region. When the grain surface was coated with \ce{LiAlO2}(LAO), the critical current density was enhanced due to a reduction in the electronic conductivity of the grain boundaries. In addition, electronic conductivity can increase nonlinearly with external potential, aggravating failure~\cite{wangNanoscaleOriginSofttoHard2025,songRevealingShortCircuitingMechanism2019}.

\begin{figure}[htbp]
    \centering
    \includegraphics[width=1\textwidth]{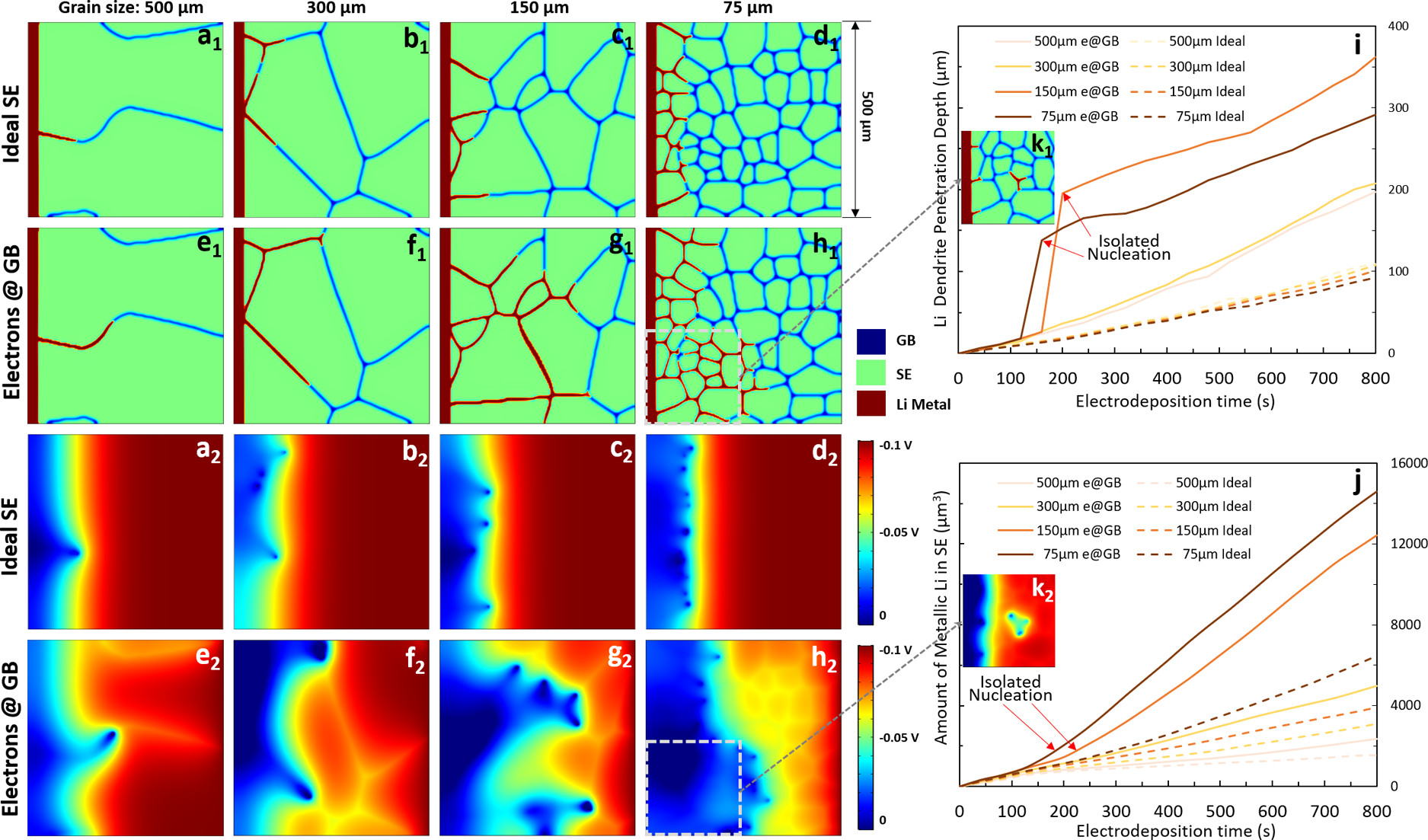}
    \caption{a1-d1, a2-d2) Simulation results of dendrite growth in ideal SE with varying grain sizes. e1-h1, e2-h2) Simulation results of dendrite growth in SE with trapped surface electrons with varying grain size. These simulations show that trapped electrons change the growth pattern and depth of dendrites. These growths increase in size and become more uneven when compared to the ideal SE.\cite{tianInterfacialElectronicProperties2019} i) Li penetration depth and j) amount of Li within SE as a function of the electrodeposition time.  Reproduced with permission from American Chemical Society.}
    \label{fig:tian}
\end{figure}
The electronic leakage mechanism has improved our understanding of soft shorts in SSBs, which are characterized by the flow of both ionic and electronic current through the SE. ~\citet{wangNanoscaleOriginSofttoHard2025} performed Li plating experiments with SSBs inside a transmission electron microscope and observed the transition from soft to hard short circuit. They found clear evidence of complete lithiation of the \ce{Li_{1.3}Al_{0.3}Ti_{1.7}(PO4)3} SE caused by the reduction of \ce{Li+} to \ce{Li^0} within the SE due to electronic leakage. This led to frequent short circuit events during Li plating, eventually followed by complete short circuit when the \ce{Li^0} domains were fully connected.
\begin{figure}[htbp]
    \centering
    \includegraphics[width=.7\textwidth]{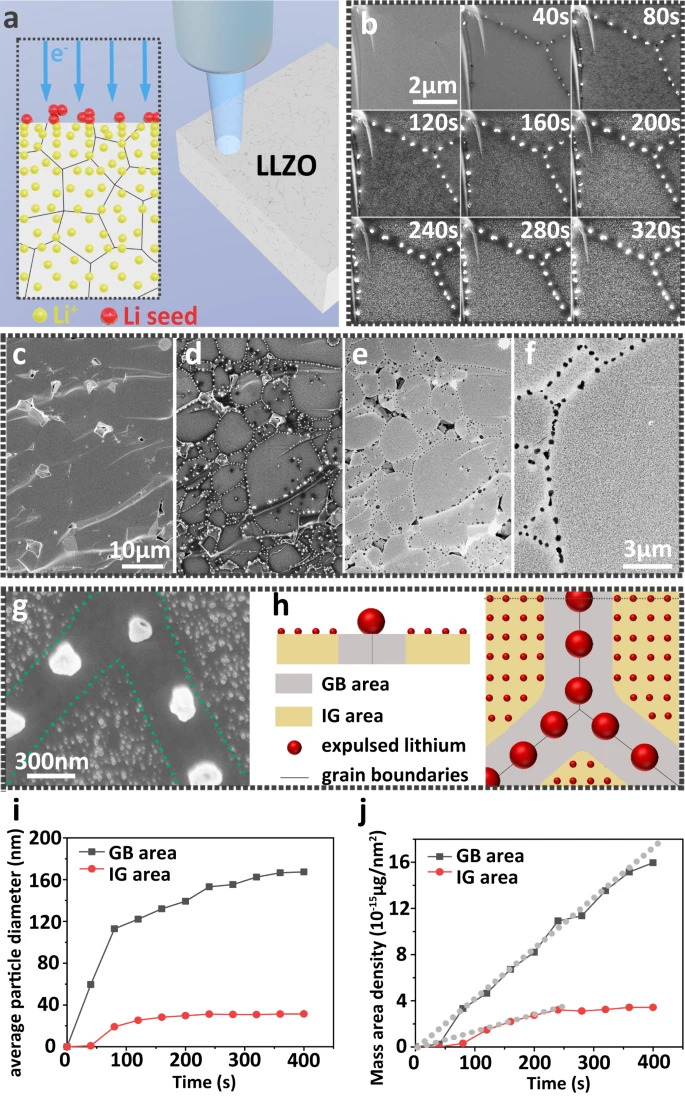}
    \caption{(a) Schematic of experimental setup with electron injection into LLZO. (b) Increase in Li particle size as time under electron injection increases. (c-d) Images of base and 400-second cross section, respectively, taken with secondary electron detection mode. (e-f) Images taken after 400 seconds with backscattering electron detector mode. (g) Enlarged image of grain boundary and inner grain (IG) areas showing that Li metal propagates into larger particles along the grain boundary compared to the inner grain. (h) Schematic representing the phenomenon occurring in g. (i-j) Graph showing that grain boundary areas are more prone to Li+ reduction than IG areas. Reproduced from Refs.~\cite{zhuUnderstandingEvolutionLithium2023} under the terms of the Creative Commons CC BY license.}
    \label{fig:zhu2023}
\end{figure}
\citet{zhuUnderstandingEvolutionLithium2023} used in situ Kelvin probe force microscopy on a symmetric LLZO cell to map the local electrical potential in the SE as shown in \autoref{fig:zhu2023}. Their measurements showed a sharp drop in potential near grain boundaries due to the accumulation of large amounts of electrons, which reduced \ce{Li+} to form dendrites. Such a drop in potential was not observed in amorphous \ce{Li3PO4}, indicating that grain boundaries strongly contributed to dendrite growth.

Atomistic and phase-field simulations are a useful tool to determine the local electronic conductivity at grain boundaries and how they affect the dendrite growth morphology. DFT calculations confirmed that the band gap considerably reduces near grain boundaries for all SE classes~\cite{xie_effects_2024}. 
\citet{tianInterfacialElectronicProperties2019} combined DFT simulations with phase-field modeling to model dendrite growth through grain boundaries. They found that electron trapping near surfaces led to faster dendrite growth and Li nucleation within the SE, as shown in \autoref{fig:tian}.

\subsection{Strategies to Mitigate Dendrite Growth}
Modifying the properties of grain boundaries is essential to prevent dendrite growth and failure, as both their mechanical and electronic properties play an important role. Common strategies employed to address this challenge are developing composite SEs; introducing a secondary phase into the grain boundaries, or eliminating the grain boundaries; and modifying the electrode-electrolyte interface. 

Composite ceramic-polymer electrolytes have widely been used to modulate both the mechanical, ionic, and electronic properties of grain boundaries to prevent dendrite growth~\cite{wangNanoscaleOriginSofttoHard2025,counihanPhantomMenaceDynamic2024,fuUniversalChemomechanicalDesign2020a}. The polymer used to infiltrate grain boundaries offers several major benefits: 1) it blocks electron transport through grain boundaries, 2) it provides mechanical resilience to accommodate volume changes, 3) it serves as an organic binder providing effective adhesion between the grains and preventing Li growth~\cite{wangNanoscaleOriginSofttoHard2025,counihanPhantomMenaceDynamic2024}. No short circuit and stable plating without soft shorts have been demonstrated with some composite SEs.

\citet{Du2024} prepared amorphous SE with a mixture of LLZO and \ce{Li3BO3}, which creates a three-dimensional continuous grain boundary network that prevents dendrite growth by blocking electron pathways. 
Grain boundary amorphization is another mitigation strategy that involves heating the SE until the grain boundary melts and then performing vitrification, wherein the SE is cooled with no grain boundary.~\cite{You2025} This creates a mechanically stronger SE and reduces Li aggregation. It reduces dendrite penetration, accompanied by a small decrease in the ionic conductivity. Another possible solution is to use doping to fill and densify the grain boundary using different additives ball-milled with the SE.~\cite{zhengGrainBoundaryModification2021} \citet{Lee2024} milled \ce{Li_2O-B_2O_3-Al_2O_3} along with \ce{Li_{6.1}Ga_{0.3}La_3Zr_2O_{12}} and sintered the solution together to improve the grain size and coat the grain boundary. They discovered that the denser pellet exhibited protection against mechanical defects, lowered electron conductivity, and improved ionic conductivity along the grain boundary.~\cite{Lee2024} Another way to densify the pellet is to use a smaller particle size in the synthesis. This allows lower energy to be used during the heating and compression of the pellet and creates a final product with fewer voids. 
Additives that reduce the electronic conductivity of grain boundaries, such as \ce{Al2O3} have been shown to suppress dendrite growth~\cite{huangEnhancedPerformanceLi2020}.
Some studies focused on modifying the electrode/electrolyte interface alone to prevent dendrite growth. ~\citet{Kim2020}  used laser sintering to produce an amorphous layer on the outer edges of LLZO. This helped prevent dendrite growth while minimizing the drawbacks of traditional amorphous SE. \citet{sastreBlockingLithiumDendrite2021} used physical layer deposition to control the properties of the coating, allowing them to tune the amount of Li in the surface layer and increasing the ionic conductivity by four orders of magnitude.

\clearpage
\section{Role of Grain Boundaries in Void Formation and Contact Loss}\label{sec:void}

SSBs exhibit failure during metal stripping at high discharge rates due to void formation and contact loss~\cite{kasemchainanCriticalStrippingCurrent2019a,lewisLinkingVoidInterphase2021}. 
Voids form when the rate of vacancy diffusion into Li metal is slower than the rate at which they are generated at the interface.~\cite{krauskopfDiffusionLimitationLithium2019} This is accompanied by interfacial delamination, void coalescence, and current constriction, accelerating dendrite formation during plating~\cite{krauskopfDiffusionLimitationLithium2019,zhaoImagingEvolutionLithiumsolid2025,limonConstrictionContactImpedance2025}.

Grain boundaries in metallic anodes and SEs critically influence void formation and interfacial stability in SSBs.  Grain boundaries within both the SE and metal anode introduce non-uniform electrochemical, mechanical, and transport properties, which can accelerate void formation due to current focusing and constriction~\cite{krauskopfPhysicochemicalConceptsLithium2020,limonConstrictionContactImpedance2025}. 
Repeated plating and stripping cycles was found to cause interfacial voids and contact inhomogeneity in LLZO, which was attributed to nonuniform Li conductivity at grains and grain boundaries\cite{tsaiLi7La3Zr2O12InterfaceModification2016}.  
Similarly, \citet{sadowski2024grain} reported that grain boundaries in argyrodite Li$_6$PS$_5$Br exhibit excess volume and structural disruption, which may serve as precursors to local softening, delamination, or contact loss.  
The Li-rich or depleted grain boundary cores can further exacerbate inhomogeneities and lead to degraded contact at high stripping rates~\cite{guAtomicscaleStudyClarifying2023,sasano2021atomistic}.

\begin{figure}[htbp]
    \centering
    \includegraphics[width=1.0\textwidth]{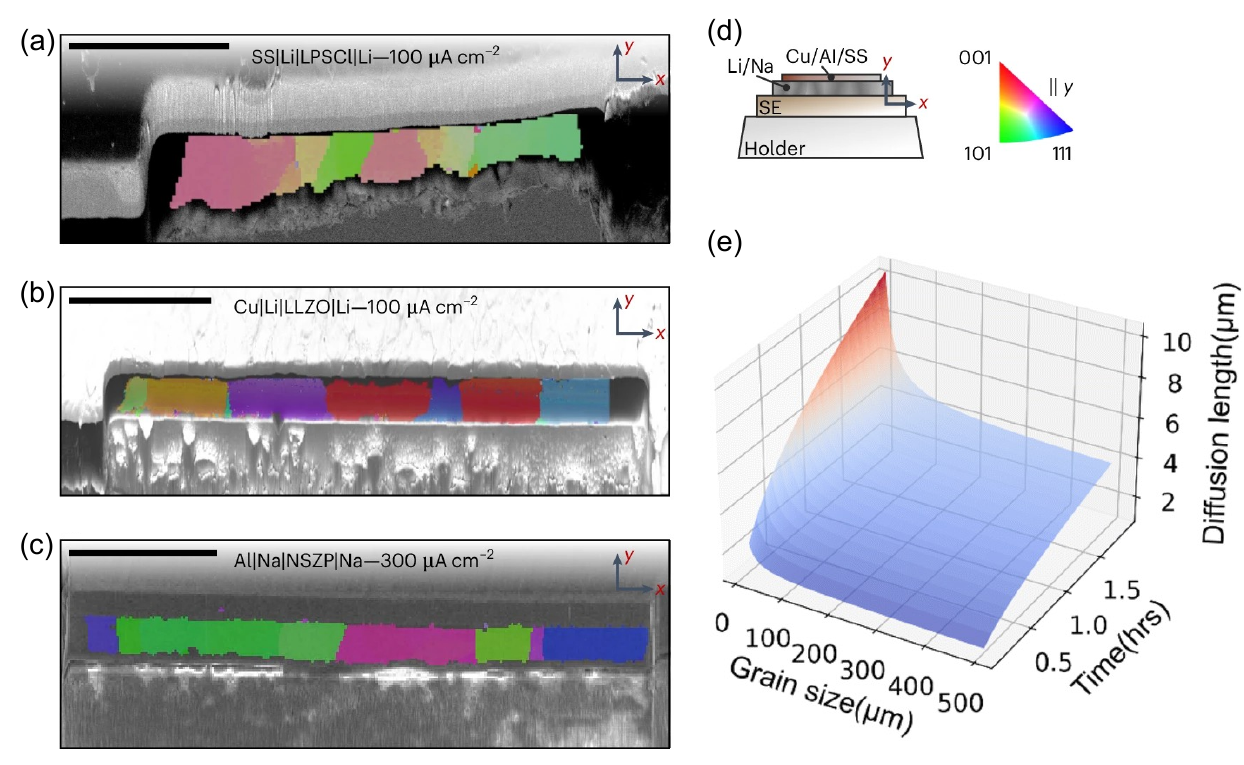}
    \caption{(a–c) IPF maps showing the grain structure of electrodeposited alkali metal layers at different current collector|Solid Electrolyte (CC|SE) interfaces, adapted from Fuchs et al.~\cite{fuchsImagingMicrostructureLithium2024}. Li was deposited in (a) SS$\mid$Li$_6$PS$_5$Cl$\mid$Li and (b) Cu$\mid$Li$_{6.5}$Ta$_{0.5}$La$_3$Zr$_{1.5}$O$_{12}$$\mid$Li cells at 100 $\mu$A cm$^{-2}$, while sodium was deposited in (c) Al$\mid$Na$_3$Zr$_2$Si$_2$P$_{0.6}$O$_{12}$$\mid$Na cells at 300 $\mu$A cm$^{-2}$. Large, columnar grains perpendicular to the interface are observed in all systems, suggesting growth governed by interfacial kinetics rather than substrate microstructure. (d) Schematic of the EBSD cross-sectional imaging geometry and IPF color key. (e) Simulated Li diffusion length as a function of grain size and time at 300 K, adapted from Yoon et al.~\cite{yoonExploitingGrainBoundary2023}, highlighting that sub-micron grains enhance mass transport via fast grain boundary diffusion. Together, these results emphasize the role of alkali metal grain morphology and grain boundary transport in stabilizing anode interfaces in SSBs. Reproduced from Refs.~\cite{fuchsImagingMicrostructureLithium2024,yoonExploitingGrainBoundary2023} under the terms of the Creative Commons CC BY license.}
    \label{fig:anode_grain}
\end{figure}
The grain boundaries and microstructure of the metal anode also critically influence the void formation. 
~\citet{fuchsImagingMicrostructureLithium2024} visualized Li and Na deposition in anode‐free cells using electron backscatter diffraction (EBSD) (\autoref{fig:anode_grain}d), revealing columnar grains with grain boundaries perpendicular to the SE interface as shown in \autoref{fig:anode_grain}a-c.  
Cross-sectional inverse pole figure (IPF) maps of Li plated at the SS/Li$_6$PS$_5$Cl (100~$\mu$A~cm$^{-2}$, 15~MPa) and Cu/Li$_{6.5}$Ta$_{0.5}$La$_3$Zr$_{1.5}$O$_{12}$ (100~$\mu$A~cm$^{-2}$, 5~MPa) interfaces, and Na plated at the Al/Na$_3$Zr$_2$Si$_2$P$_{0.6}$O$_{12}$ (300~$\mu$A~cm$^{-2}$, 3~MPa) interface, show that the average grain size of each electrodeposited metal film is quite large (10–150~$\mu$m), with grain boundaries oriented vertically across the thickness of the metal layer.  
These grain structures were found to be independent of current collector or electrolyte substrate microstructure.  
Furthermore, void formation upon stripping was shown to initiate within the grains rather than at grain boundaries, emphasizing the latter as efficient diffusion conduits for vacancies.  
Yoon et al.'s~\cite{yoonExploitingGrainBoundary2023} atomistic simulations could provide a partial explanation for this phenomenon. They demonstrated that diffusion along Li grain boundaries is three to six orders of magnitude faster than in the bulk.  
Their simulations showed that reducing Li grain size to the sub-micron regime significantly increases the effective diffusivity of the anode, allowing for extended stripping times before interfacial contact loss as shown in 
~\autoref{fig:anode_grain}e.

\begin{figure}[htbp]
    \centering
    \includegraphics[width=0.95\textwidth]{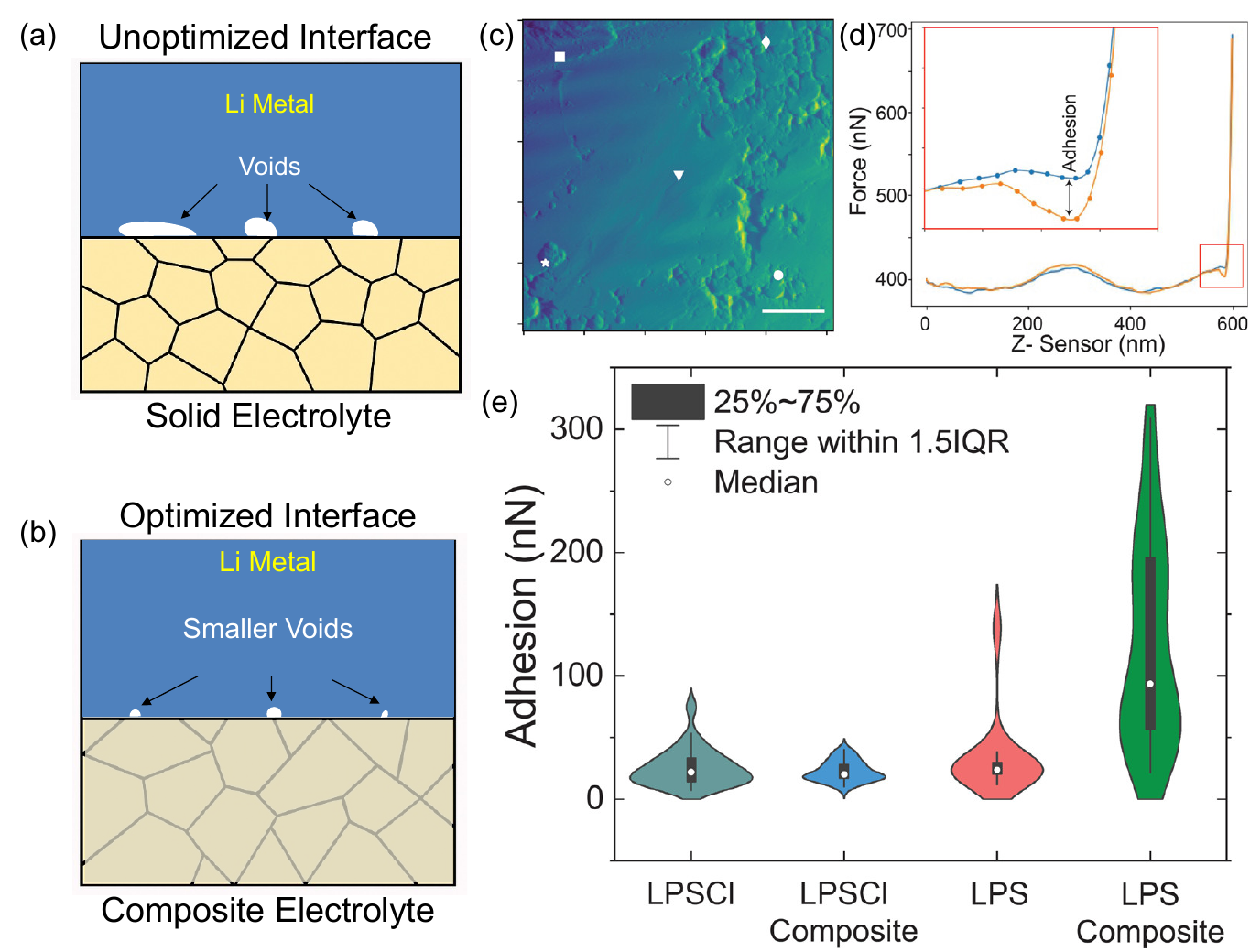}
    \caption{Schematic illustration and AFM data highlighting the role of grain boundaries in void formation and adhesion in SSBs.  
    (a) Interface schematics showing void formation at the Li metal|SE interface and (b) improved conformal contact at the Li metal|composite electrolyte interface, where polymer‐rich phases improve mechanical coupling and suppress voids.  
    (c-e) AFM-based force–distance mapping of local adhesion in LPS and LPSCl SEs and their polymer composites~\cite{dixitDifferencesInterfacialMechanical2022}.  
    (c) AFM topography with predefined measurement points.  
    (d) Example force–distance curve; adhesion is obtained from the force difference between retrace and trace at the Z-sensor position corresponding to the minimum of the retrace curve.  
    (e) Violin plots: all samples show bimodal adhesion; bare pellets have medians of $\sim$22~nN (LPSCl) and $\sim$24~nN (LPS). The LPS composite shows a much higher median ($\sim$94~nN) with a broader spread, whereas the LPSCl composite remains lower ($\sim$20$-$30~nN) with smaller variance.  
    Reproduced with permission from Ref.~\cite{dixitDifferencesInterfacialMechanical2022}. Copyright \copyright~2022, American Chemical Society.}
    \label{fig:dixit_adhesion}
\end{figure}

Void formation is also influenced by the interfacial adhesion of the metal anode and the SE.  Bare ceramic pellet surfaces have weak adhesion towards Li metal, while polymer-ceramic composites provide stronger adhesion~\cite{dixitDifferencesInterfacialMechanical2022}. This results in smaller-sized voids for composite SEs as compared in \autoref{fig:dixit_adhesion}a and b.
~\citet{dixitDifferencesInterfacialMechanical2022} mapped nanoscale adhesion in thiophosphate (LPS) and argyrodite (LPSCl) SEs and their polymer composites using AFM force–distance curves.  
As shown in ~\autoref{fig:dixit_adhesion}c, AFM topography was recorded with predefined adhesion measurement points; local adhesion was then extracted from the force difference between the retrace and trace curves at the Z-sensor position corresponding to the minimum of the retrace curve (\autoref{fig:dixit_adhesion}d).  
As shown in ~\autoref{fig:dixit_adhesion}e, bare pellets exhibited bimodal adhesion distributions ($\sim$20–24~nN), attributed to grains and grain boundaries, while polymer-rich LPS composites showed a higher and broader median adhesion ($\sim$94~nN), maintaining intimate interfacial contact during cycling.
Regions with low adhesion often correlated with grain boundaries that are prone to interfacial delamination, initiating voids, and accelerating contact loss, whereas mechanically compliant or polymer-modified surfaces prevent contact delamination.

These studies reveal that while grain boundaries facilitate faster vacancy diffusion in the metal anode, they also present a mechanical weakness in the SE due to reduced interfacial adhesion.
Optimizing ASSB performance requires dual engineering: 1) finer grains to exploit grain boundary-mediated Li diffusion and mitigate void nucleation, and 2) interfaces with high, uniform adhesion to prevent contact loss.  
Combining microstructural control with mechanically compliant interlayers offers a robust path toward void-free, dendrite-resistant, and low stack pressure SSBs.

\clearpage
\section{Comparison of Solid Electrolyte Classes}\label{sec:differences}
The differences in grain boundary properties between SEs belonging to different classes, such as oxides,  halides, sulfides, and antiperovskites, are sufficient to influence the performance of SSBs. Here, we highlight the key differences between these SE classes in terms of their ionic transport, processing, chemomechanical, and electronic properties.

\subsection{Ionic Transport}
\begin{figure}[htbp]
    \centering
    \includegraphics[width=1.0\textwidth]{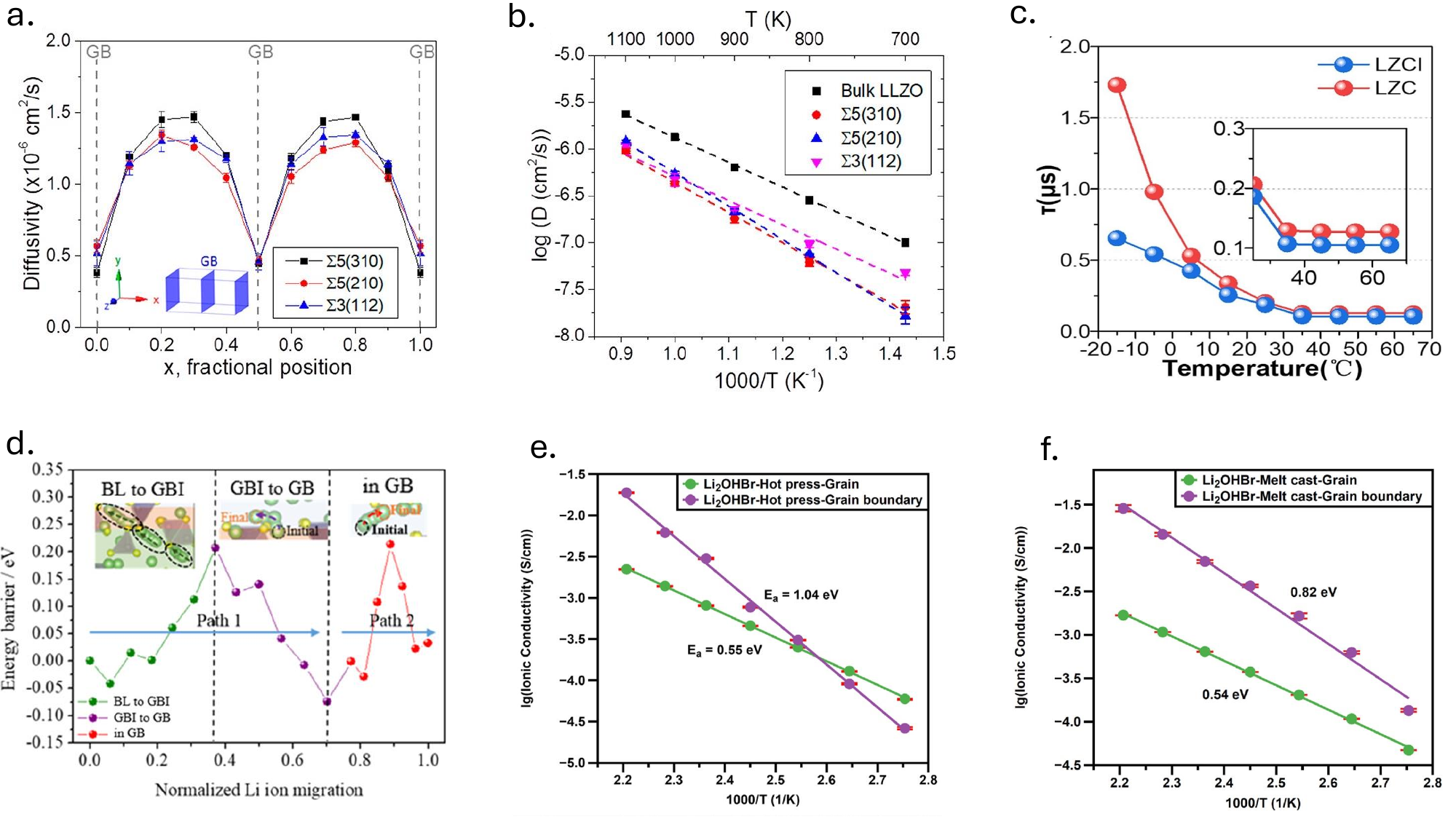}
    \caption{ (a) Variation of total Li-ion diffusivity for three grain boundary cells at 1000K with position perpendicular to the plane of grain boundary (b) Arrhenius plots of Li-ion diffusivity within the grain boundary regions across a temperature span of 700–1100 K, compared to the bulk material.\cite{yu_grain_2017} Reproduced with permission from Ref.\cite{yu_grain_2017}. Copyright \textcopyright 2017 American Chemical Society. (c) Distribution of relaxation times (DRT) curves showing the distinct time constants for bulk and grain boundary areas in Li$_2$ZrCl$_6$ and Li$_2$ZrCl$_5$I across different temperatures. Reproduced with permission from Ref.\cite{wang_superionic_2025}. Copyright \textcopyright 2025 Elsevier B.V. (d)Energy barriers of Li migration as variations of the regions for Li$_{10}$GeP$_2$S$_{12}$. Here GBI stands for Grain-bulk interface, BL means bulk like, and GB refers to grain boundary.
    Reproduced with permission from Ref.\cite{hwang_electrochemical_2023}.Copyright \textcopyright~2023 American Chemical Society. (e) Temperature-dependent ionic conductivity for hot-pressed Li$_2$OHBr anti-perovskite-type SE and (f) Temperature-dependent ionic conductivity for melt-cast Li$_2$OHBr anti-perovskite type SE. Reproduced with permission from Ref.\cite{zheng_differentiating_2022}. }
             
    \label{fig:nav-1}
\end{figure}

The impact of grain boundaries on ionic conductivity varies significantly across different classes of SEs. 
The first report on the discovery of LLZO found significantly high grain boundary resistance, amounting to 40-50\% of the total resistance~\cite{muruganFastLithiumIon2007}. 
Later studies found that grain boundaries contribute a moderate resistance on Li-ion transport\cite{yu_grain_2017}. Specific boundaries, like $\Sigma5$, can slow down Li diffusivity by two orders of magnitude than the bulk, with activation energies increasing to about 0.71 eV. Other boundaries, such as $\Sigma3$, retain roughly half of the bulk conductivity and demonstrate pronounced anisotropic transport. These discrepancies can be understood from ~\autoref{fig:nav-1}a and b.  In ~\autoref{fig:nav-1}a, the plotted diffusivity appears to peak in the bulk but significantly decreases near the grain boundary. Further, the Arrhenius plots in ~\autoref{fig:nav-1}b show that the diffusivity is nearly halved near grain boundaries compared to the bulk.
Another oxide SE, LLTO, is known to have a high grain boundary resistance, which has been the subject of numerous investigations~\cite{guAtomicscaleStudyClarifying2023,sasano2021atomistic}.

The situation is quite different for halide SEs, where the grain boundaries typically do not hinder Li-ion movement as much~\cite{asano_solid_2018}. In \ce{Li3YCl6} and \ce{Li3YBr6}, weak Li--halogen interactions, large polarizable anions, and intrinsic cation vacancies enable fast, isotropic Li$^+$ transport. Their high mechanical deformability may also help in minimizing grain boundary resistance and preserving bulk-like conductivity.\cite{asano_solid_2018}. Even further amorphization, as observed in iodide doped Li$_2$ZrCl$_6$ (Li$_2$ZrCl$_5$I) can reduce grain boundary resistance. ~\autoref{fig:nav-1}c demonstrates that Li$_2$ZrCl$_5$I has consistently shorter and more temperature-stable relaxation times than Li$_2$ZrCl$_6$, suggesting improved ionic conductivity\cite{wang_superionic_2025}.

In sulfide electrolytes, softer and more amorphous or glassy structures often promote better interparticle contact and more efficient ion transport, especially in grain boundary-free systems like xLi$_2$S–(100-x)P$_2$S$_5$.\cite{lau_sulfide_2018,sakuda_sulfide_2013} For example, $\beta$-Li$_3$PS$_4$ can exhibit enhanced conductivity at grain boundaries and amorphous/crystal interfaces, with values reaching 1 mS/cm that is one or two orders of magnitude higher than the bulk\cite{jalem_lithium_2023}. More crystalline sulfides such as \ce{Li10GeP2S12}(LGPS), display higher grain boundary resistance as seen from energy barriers for Li transport near grain boundaries in \autoref{fig:nav-1}d. \cite{hwang_electrochemical_2023}. In argyrodites, the grain boundary conductivity is sensitive to local composition, i.e., halide/S site exchange~\cite{sadowski2024grain}. 

\begin{figure}[htbp]
    \centering
    \includegraphics[width=1.0\textwidth]{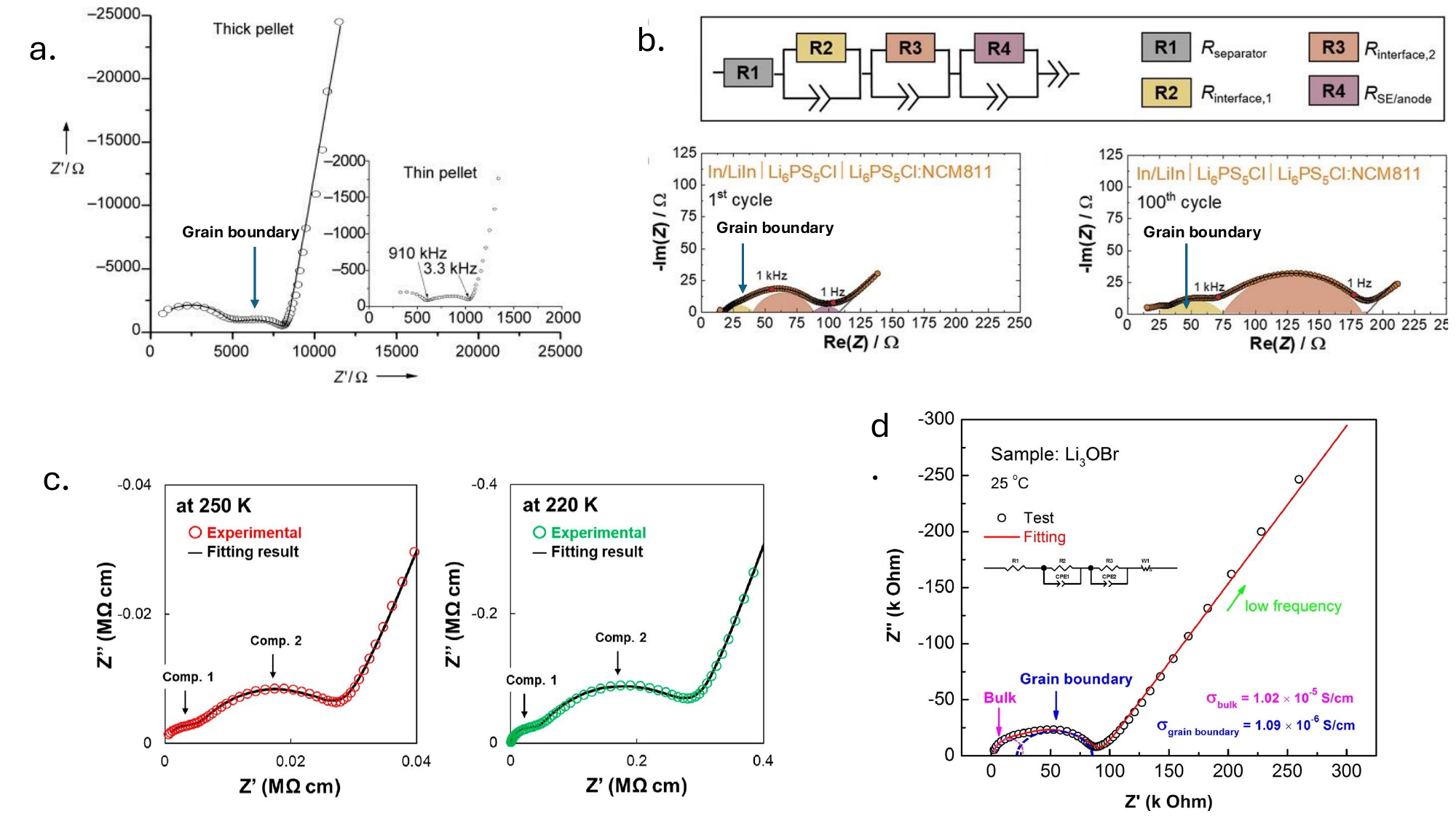}
    \caption{ EIS plots of (a) LLZO where the second semicircle represents the grain boundary resistance, Reproduced with permission from Ref.\cite{muruganFastLithiumIon2007}. Copyright \textcopyright 2007 Wiley-VCH Verlag GmbH \& Co. KGaA, Weinheim (b) Plots of the impedance data after the 1st and 100th cycle for halide type electrolytes are shown along with the equivalent circuit used for fitting. Reproduced from \citet{rosenbach2023visualizing} under the terms of the Creative Commons Attribution License (c) EIS measurements for Li$_{7-x}$PS$_{6-x}$Cl$_x$ sulphide electrolyte (LPSCl, $x \sim 1$), conducted over a temperature interval of 180–250~K. Comp 1 indicates to the bulk and component 2 is for grain boundary, Reproduced with permission from Ref.\cite{morino2023high}. Copyright \textcopyright 2023 American Chemical Society
(d)The EIS plot of Li$_3$OBr at $25\,^\circ\mathrm{C}$     Reproduced with permission from Ref.\cite{li_reaction_2016}.Copyright \textcopyright~2015 Elsevier B.V. }

    \label{fig:nav-4}
\end{figure}
In antiperovskite SEs, in the case of Li$_3$OCl, specific low-energy grain boundary structures like $\Sigma3$ may restrict ionic movement in contrast to other grain boundaries like $\Sigma5$\cite{dawson_atomic-scale_2018, chen_insights_2018}. Grain boundaries tend to raise activation energies (typically 0.40–0.56 eV, compared to 0.29 eV in the bulk), resulting in a substantial reduction in conductivity.\cite{dawson_atomic-scale_2018}. \citet{zheng_differentiating_2022} obtained higher grain boundary conductivity but also higher activation energy in \ce{Li2OHBr} SE. However, as illustrated in \autoref{fig:nav-1}e and f, the melt-case sample had a lower activation energy for grain boundary transport compared to the hot-pressed sample. Another study shows that Li-vacancy segregation at grain boundaries in polycrystalline \ce{Li3OCl} can significantly reduce ionic diffusivity, but this negative effect decreases as grain size increases and grain boundaries become more coherent\cite{shen_revealing_2020}. In Li$_2$OHCl, high Li-ion conductivity arises from the synergistic effect of cation vacancies and dynamic OH$^-$ anion rotation, especially in the cubic phase above 311 K.\cite{wang_dynamics_2020}
\autoref{fig:nav-4} shows the impedance spectrum of four different classes of SEs: a) oxide, b) halide, c) sulfide, and d) antiperovskite, clearly showing the semicircles associated with bulk and grain boundary resistance.

\subsection{Processing and Chemomechanical Properties}
As grain boundaries have a different influence on each type of SE, the processing strategy used must be tailored for the  SE under consideration. The processing techniques and steps, such as hot pressing, synthesis routes, spark plasma sintering, two-step sintering, and doping have a major impact on grain boundary structure and integrity\cite{hikima2022mechanical}. 
Higher density of SEs from optimized processing improves grain boundary cohesion, while insufficient densification or residual impurities can weaken boundaries and reduce mechanical strength. Synthesis methods that yield smaller particles at grain boundaries increase the risk of stress concentration and softening. Additive manufacturing and 3D printing techniques have shown promise for spatially resolved grain boundary engineering through controlled mesoscale architectures~\cite{chen2020manufacturing}. However, challenges remain in scaling and interfacial stability. A coextrusion-based strategy has been employed to achieve macroscale control over interfaces, enabling tunable ionic conductivities and underscoring the critical role of interfacial phase continuity in optimizing performance.~\cite{dixit2019scalable}. These insights into processing and interface control set the stage for understanding how such factors directly affect the mechanical robustness of different SE materials.

\begin{figure}[htbp]
    \centering
    \includegraphics[width=0.5\textwidth]{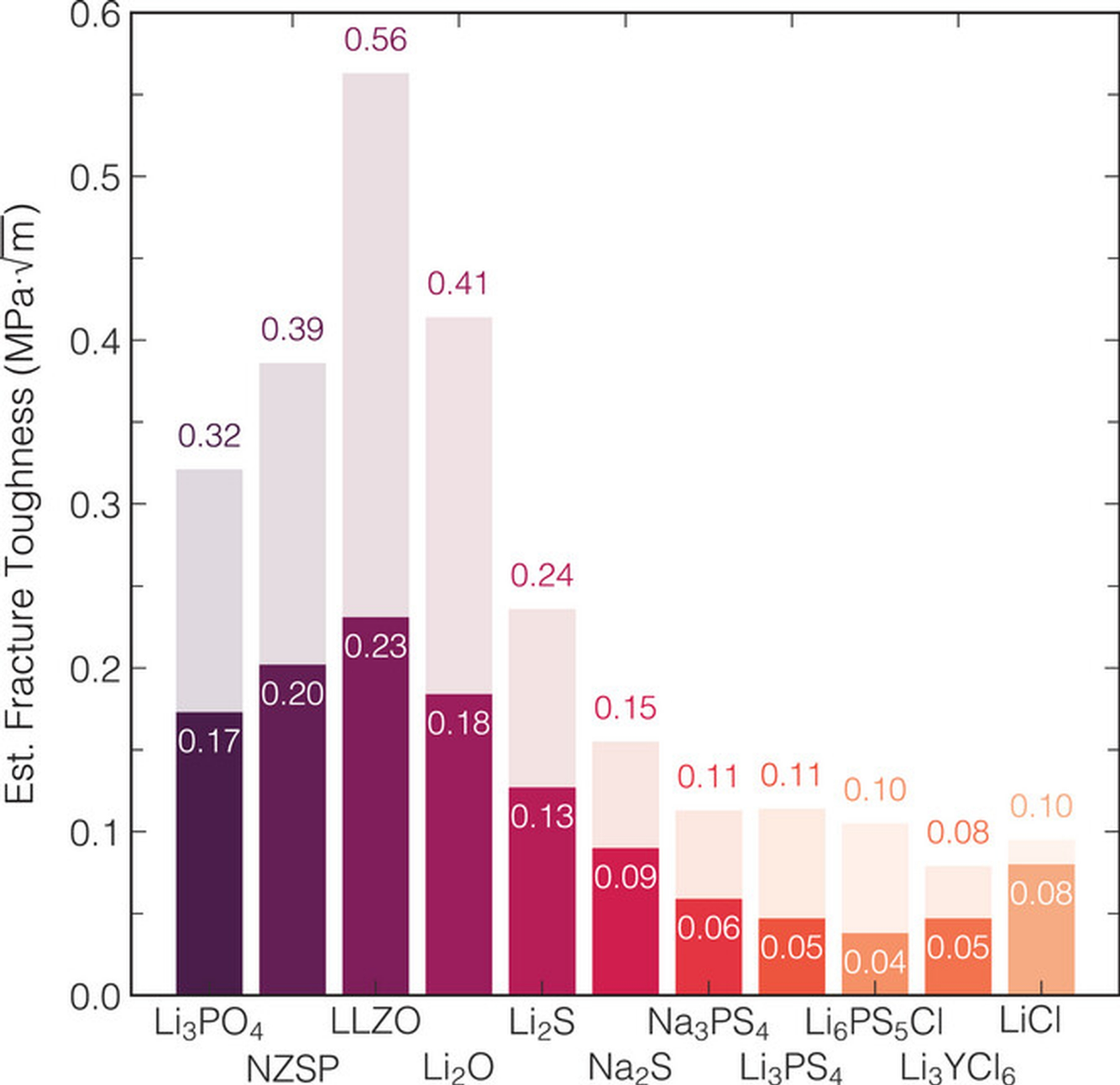}
    \caption{Fracture toughness values (\(K_{\mathrm{Ic}}\)) for various SEs for the bulk (lighter) and grain boundary regions (darker) obtained from first-principles calculations.
    LLZO corresponds to \(\mathrm{Li_7La_3Zr_2O_{12}}\) and NZSP to \(\mathrm{Na_3Zr_2Si_2PO_{12}}\) Reproduced from ~\citet{xie_effects_2024} under the terms of the Creative Commons Attribution License.
 }
    \label{fig:nav-5}
\end{figure}

\autoref{fig:nav-5} shows that oxide SEs exhibit much higher fracture toughness than sulfides and chlorides, with LLZO displaying the highest values for both grain boundaries and bulk\cite{xie_effects_2024}. In contrast, Li$_6$PS$_5$Cl has the lowest grain boundary fracture toughness, while Li$_3$YCl$_6$ has the lowest bulk value. Generally, the bulk fracture toughness is approximately double that of grain boundaries.
The following sections examine processing methods and their influence on grain boundaries in oxide, halide, sulfide, and antiperovskite SEs.

In case of oxide SEs such as LLZO, Σ5(310) boundaries and Li rich phases can embrittle the SE and decrease ionic conductivity\cite{monismith_grain-boundary_2022, cojocaru-miredin_quantifying_2022}.
~\citet{huangEnhancedPerformanceLi2020} investigated how starting particle size (micron vs. nano) and an Al$_2$O$_3$ (AO) sintering additive affect the properties of 
\ce{Li_{6.4}La3Zr_{1.4}Ta_{0.6}O12} (LLZTO) SEs. They fabricated four distinct pellets: micron-sized with and without the additive (M-LLZTO, M-LLZTO-AO) and nano sized with and without the additive (N-LLZTO, N-LLZTO-AO).
\begin{figure}[htbp]
    \centering
    \includegraphics[width=0.9\textwidth]{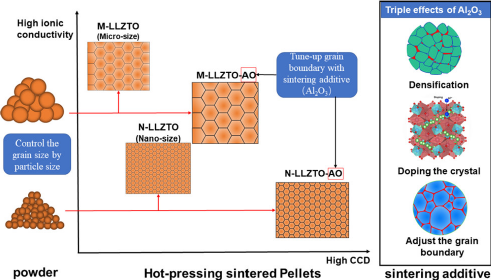}
    \caption{Processing–microstructure–property relationship in LLZTO. Particle size controls grain size, and Al$_2$O$_3$ sintering additives tune the grain boundary via densification, lattice doping, and LiAlO$_2$ interphase formation. Adapted with permission from Huang et al.~\cite{huangEnhancedPerformanceLi2020}. Copyright \textcopyright~2020 American Chemical Society.}
    \label{fig:huang_toc}
\end{figure}
\autoref{fig:huang_toc} shows that particle size controls grain size, with nano powders producing fine grains and micro powders giving coarse grains. \ce{Al2O3} additive further modifies grain boundaries through densification and interphase formation, enhancing critical current density while maintaining ionic conductivity.
\begin{table}[htbp]
    \centering
    \caption{Relative density, total ionic conductivity, and activation energy ($E_a$) of LLZTO samples with different particle sizes and Al$_2$O$_3$ sintering additives. Data reproduced with permission from Huang et al., \textit{ACS Appl. Mater. Interfaces} \textbf{2020}, \textit{12}, 56118–56125. Copyright (2020) American Chemical Society.\cite{huangEnhancedPerformanceLi2020}}
    \label{tab:huang_table1}
    \begin{tabular}{lccc}
        \hline
        \textbf{Samples} & \textbf{Relative density (\%)} & $\boldsymbol{\sigma_\text{total}}$ (S\,cm$^{-1}$) & $\boldsymbol{E_a}$ (eV) \\
        \hline
        N-LLZTO-AO & 97.3 & $3.60 \times 10^{-4}$ & 0.36 \\
        N-LLZTO    & 96.5 & $5.15 \times 10^{-4}$ & 0.33 \\
        M-LLZTO-AO & 98.5 & $6.42 \times 10^{-4}$ & 0.34 \\
        M-LLZTO    & 98.0 & $1.01 \times 10^{-3}$ & 0.29 \\
        \hline
    \end{tabular}
\end{table}
As summarized in Table~\ref{tab:huang_table1}, the sample made from M-LLZTO achieved the highest total ionic conductivity ($1.01 \times 10^{-3}$~S~cm$^{-1}$), due to larger grains and fewer grain boundaries. In contrast, N-LLZTO-AO demonstrated the best performance in suppressing dendrite growth and had the highest critical current density, despite its lower conductivity.

\begin{figure}[htbp]
    \centering
    \includegraphics[width=1.0\textwidth]{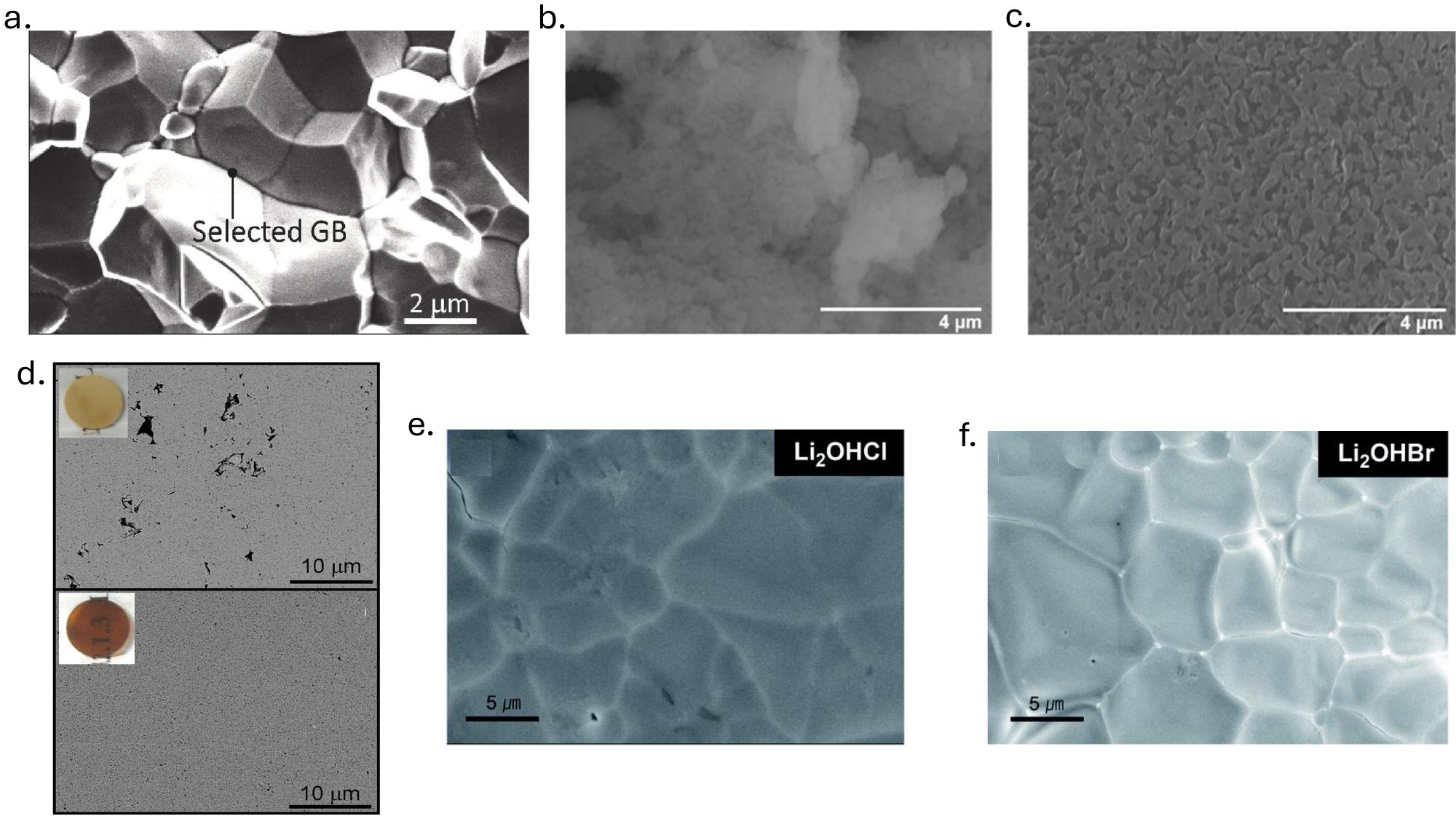}
    \caption{(a)SEM image of LLZO highlighting the grain boundary. Reproduced with permission from Ref.~\cite{cojocaru-miredin_quantifying_2022}. Copyright \textcopyright2022 Elsevier B.V.SEM image of (b) cold-pressed LYBC-BM(ball-milled) and (c) hot-pressed LYBC-HP. Reproduced with permission from Ref.~\cite{liu_high_2021}. Copyright \textcopyright~2020 American Chemical Society. (d)Cross-sectional SEM images of 80Li$_2$S·20P$_2$S$_5$ glassy SEs pressed at 360\,MPa, shown at 25\,$^\circ$C and 200\,$^\circ$C. Reproduced from \citet{sakuda_sulfide_2013} under the terms of the Creative Commons Attribution License (e)Grain structure of a Li$_2$OHCl pellet. Reproduced from ~\citet{lee2022li} under the terms of the Creative Commons Attribution License (f) Grain structure of Li$_2$OHBr pallet.
 }
    \label{fig:nav-2}
\end{figure}

More broadly, densifying the material through high-temperature sintering (typically 1100 – 1220$^\circ$C) is generally required to lower grain boundary resistance in oxide SEs. For example, Al-doped LLZO can achieve optimal conductivity of $2.1 \times 10^{-4}$ S/cm at 1210$^\circ$C\cite{xue_effect_2018}. However, \citet{Lee2024} observed a contrasting result for \ce{Li_{6.1}Ga_{0.3}La_3Zr2O12}:  high-temperature sintering increased the grain boundary resistance.  Adding $\mathrm{Li_2O\text{--}B_2O_3\text{--}Al_2O_3}$ sintering aids enabled liquid-phase sintering, removed grain boundary voids, improved densification, and lowered grain boundary resistance. Doping with elements such as Al and Gd can further decrease grain boundary resistance by creating Li+ vacancies and densifying the material, thereby enhancing ionic conductivity\cite{ahmad2025reducing}. \citet{kaneko_investigation_2023} investigated that Sr doping can reduce grain boundary resistance and enhance Li ion conductivity in garnet LLZO. Incorporating specialized interfacial layers, like Ga-Au at the grain boundary, has shown promise in enhancing the ionic conductivity of LLZO.\cite{kim2024stabilization}  
\autoref{fig:nav-2}a shows an SEM image of LLZO showing the grain boundary. By using fluorination to deposit fluoride phases at the surface and grain boundaries of LLZO, \citet{wang2024key} were able to block electron conduction and prevent dendrite formation. \citet{neumann2021effect} also found that optimizing processing techniques to control grain size, porosity, and phase purity is essential for improving ionic transport and preventing dendrites in garnet SEs. 
Characterization techniques such as thermal grooving\cite{sato2020mapping} and cryo-SIMS imaging\cite{D3TA05012B} are useful to visualize grain boundaries in SEs. 

In halide SEs, the processing technique used and the resulting microstructure of the SEs play a similarly complex role. For example, in amorphous xLi$_2$S–(100–x)LiI SEs, increasing the amorphous content (up to x = 70) decreases grain boundary resistance, resulting in better conductivity.\cite{di_effect_2023} But adding too much Li$_2$S can bring grain boundary resistance back, reducing the overall ionic transport. Mixed-halide systems like Li$_3$YBr$_3$Cl$_3$ (LYBC), especially when hot pressed at moderate temperatures ($\sim$170$^\circ$C), display exceptional conductivities (up to 7.2 mS/cm) due to partial grain boundary melting and enhanced grain boundary contact\cite{liu_high_2021}. SEM images (\autoref{fig:nav-2}b and c)\cite{liu_high_2021} show that hot pressing LYBC results in larger, fused particles (approximately 500--800 nm) and a denser microstructure, compared to cold-pressed samples, which exhibit smaller particles (approximately 200--300 nm) and micrometer sized voids.
Approaches like slurry coating, ethanol-based processing with ethyl cellulose binders, and mechanical treatments such as ball-milling are some promising methods to create more favorable microstructures\cite{zhang_all-solid-state_2018,wang_halide_2021}. Targeted doping is another powerful tool. For example, introducing a low-temperature molten salt (LiCl+1.33AlCl3) doping at the particle boundaries of Li$_3$GaF$_6$ forms more conductive Cl-doped Li$_3$GaF$_{6-x}$Cl$_x$, which reduces Li migration barriers and boosts ionic conductivity\cite{doi:10.1021/acsnano.4c12399}.

Sulfide-based SEs have the advantage of room temperature sintering compared to oxide SEs\cite{perrenot2024room}.  They benefit from cold pressing and careful binder optimization, which help maintain close contact between particles and keep grain boundary impedance low\cite{liu_priority_2023}. In Na-based SEs, structural continuity in materials like Na$_3$PS$_4$ ensures minimal grain boundary resistance, unlike their oxide counterparts such as Na$_3$PO$_4$, where over-coordination at grain boundaries can hinder transport\cite{dawson_toward_2019}. \autoref{fig:nav-2}d presents a cross-sectional SEM image of the 80Li$_2$S-20P$_2$S$_5$ glassy SE compressed at 360 MPa and 25$^\circ$C. The image reveals a dense microstructure with minimal visible grain boundaries, suggesting that the particles have grown together and increased in size.\cite{sakuda_sulfide_2013}
Coating grains with protective shells such as LiNbO$_3$ showed promise to improve solid-solid ionic contact and mitigate unwanted reactions, thereby suppressing the space-charge effect\cite{wang_all-solid-state_2022}. Infusing the grain boundaries of SEs like LGPS with phases like \ce{Li3PO4} can improve ionic conductivity, prevent intergranular cracking, and unfavorable phase transformation. Furthermore, embedding thin amorphous layers between grains has proven to be an effective way to maintain high ionic conductivity across the microstructure\cite{ma_recent_2018}.

Hot-pressed sulfide SE pellets show higher Meyer hardness ($0.66\,\text{GPa}$) and elastic modulus ($21\,\text{GPa}$) than cold-pressed pellets ($0.17\,\text{GPa}$, $9.5\,\text{GPa}$), due to higher density and lower porosity. \cite{hikima2022mechanical} Mechanical milling slightly improves these properties (0.73 GPa, 23 GPa) compared to liquid-phase synthesis, which results in smaller particles, more grain boundaries, and possible organic residues that soften the material. Although their softness aids battery processing, weak grain boundaries can encourage Li dendrite growth, emphasizing the need for careful grain boundary engineering.

For antiperovskite SEs, techniques such as melt casting, pulsed laser deposition, and the deliberate layering of different phases, including the integration of \ce{Li7O2Br3} have all shown promising results to optimize grain boundary contact and promote more efficient ion transport\cite{zheng_differentiating_2022, dixit_tailoring_2023, deng_span_2022, zhu_enhanced_2016}.
\autoref{fig:nav-2}e and f present two different antiperovskite SEs with visible grain boundaries\cite{lee2022li}.  High-temperature treatments and spark plasma sintering are especially effective in Na-based anti-perovskites, where they facilitate grain growth and further reduce grain boundary impedance\cite{yang_research_2025}.

\begin{figure}[htbp]
    \centering
    \includegraphics[width=0.9\textwidth]{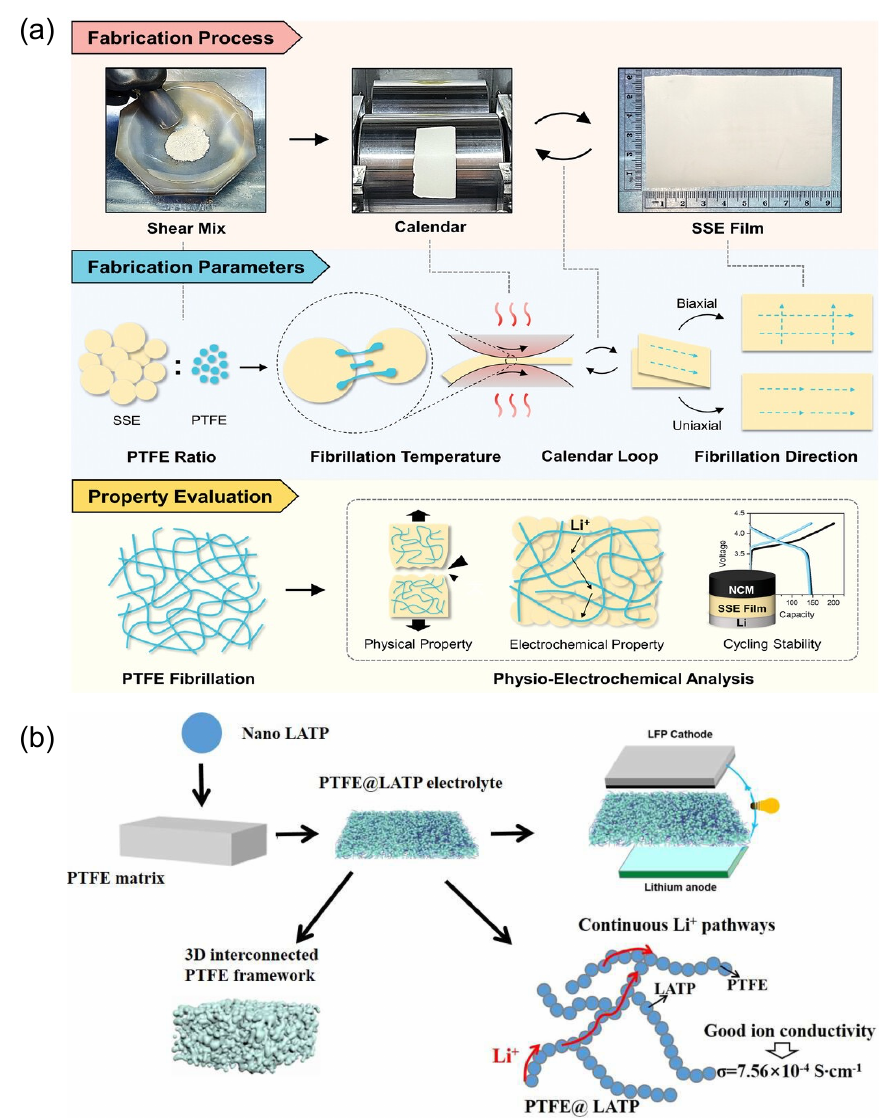}
    \caption{(a) Schematic illustration of the dry-processed SE film fabrication, highlighting the overall process flow, key fabrication parameters, and subsequent property evaluation. Reproduced with permission from ~\citet{leePhysioElectrochemicallyDurableDryProcessed2023}. Copyright \textcopyright~2023 Wiley-VCH.
    (b) Schematic of the PTFE@LATP composite electrolyte: nano-LATP is incorporated into a PTFE matrix to form a 3D interconnected PTFE framework and PTFE@LATP membrane, which in an LFP\,$|$\,Li cell provides continuous Li$^+$ pathways and improved interfacial contact, yielding $\sigma \approx 7.56\times10^{-4}\,\mathrm{S\,cm^{-1}}$. Reproduced with permission from~\citet{wang2023constructing}. Copyright \textcopyright~2023 Elsevier.
    }
    \label{fig:physio}
\end{figure}

Dry processing of SE films has recently emerged as a scalable, energy-efficient approach for SSB  manufacturing as an alternative to conventional wet and sintering-based routes. \citet{leePhysioElectrochemicallyDurableDryProcessed2023} systematically studied dry-processed \ce{Li6PS5Cl} (LPSCl) films obtained using PTFE-binder fibrillation. Beyond its role as a mechanical binder, PTFE has also been shown to significantly reduce grain boundary impedance by improving particle-particle contact and filling interfacial voids, as demonstrated in NASICON-type LATP composites by \citet{wang2023constructing}. As illustrated in \autoref{fig:physio}a, the workflow shear-mixes LPSCl powder with PTFE to form a dough, followed by calendaring and repeated folding and re-calendering (“calendar loops”) to increase PTFE fibrillation. Four fabrication parameters govern the film properties: PTFE loading ratio, fibrillation temperature, number of calendar loops, and fibrillation direction; these together dictate mechanical properties, ionic conductivity, and overall film impedance. A clear trade-off was observed: highly fibrillated films achieve excellent mechanical strength but are prone to PTFE reduction and current leakage, whereas low-fibrillation films are electrochemically more stable but mechanically weaker. Practically, optimizing fibrillation to maximize interparticle contacts while minimizing continuous, reduction-prone binder networks enables scalable manufacturing with lower film impedance and improved cycling stability. The dry route also avoids solvent-induced SE degradation and eliminates the energy-intensive drying step that constitutes roughly 47\% of total energy in wet processing, making it attractive for industrial-scale production.
Building on the benefits of PTFE-assisted processing, \citet{wang2023constructing} demonstrated that embedding nano-LATP particles into a three-dimensional interconnected PTFE framework (\autoref{fig:physio}b) not only enhances mechanical integrity and electrode–electrolyte contact but also significantly reduces grain boundary impedance. 
The PTFE network in the composite fills these voids and disperses LATP particles uniformly, creating continuous \ce{Li+} pathways and improving percolation. 
PTFE acts as a flexible, chemically stable binder that bridges particles, suppresses microcracks, and maintains intimate interparticle contact, thereby lowering both grain boundary and electrode–electrolyte impedance. 
Together, the insights from \citet{leePhysioElectrochemicallyDurableDryProcessed2023} and \citet{wang2023constructing} highlight PTFE’s dual role in scalable dry manufacturing.

Taken together, these studies underscore the central role of processing and manufacturing in designing grain boundary networks that optimize ionic transport and mechanical stability. Rational control over grain size, sintering environment, and interfacial chemistry offers a tangible pathway toward high-performance, dendrite-resistant SEs.

\subsection{Electronic Conductivity and Dendrite Suppression}
Grain boundaries influence dendrite formation due to higher local electronic conductivity. The electronic conductivity of grain boundaries greatly varies by SE type.
\begin{figure}[htbp]
    \centering
    \includegraphics[width=0.8\textwidth]{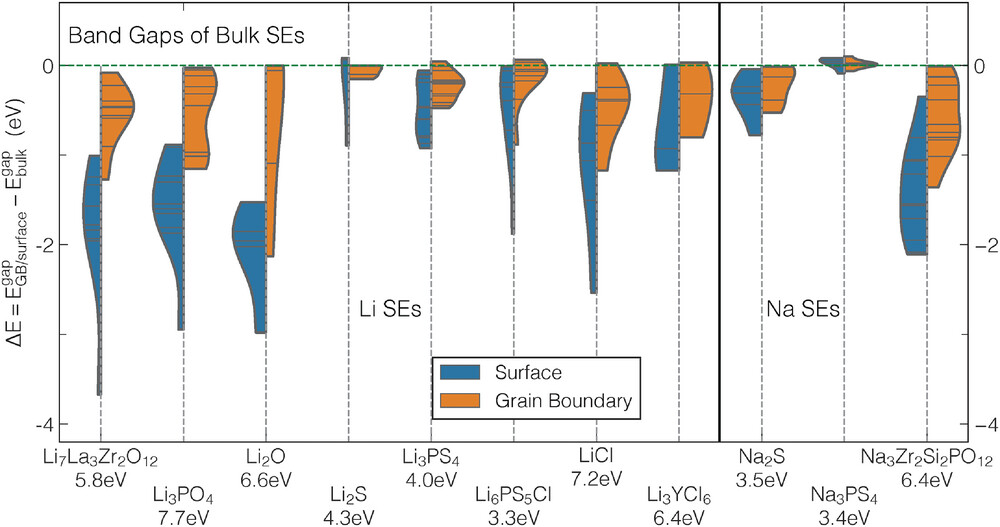}
    \caption{The variation in predicted bandgaps for grain boundaries and surfaces of SEs with their bulk counterparts. Orange shapes show the distribution of bandgap changes for grain boundaries, while blue shapes show the distribution for surfaces. The width of the violin shapes indicates the probability of finding stable models with that specific bandgap change.Figure reproduced from \citet{xie_effects_2024} under the terms of the Creative Commons Attribution License.
 }
    \label{fig:electro-nav}
\end{figure}
\autoref{fig:electro-nav} illustrates the bandgap reduction at the grain boundaries of various SEs, compared to their bulk values\cite{xie_effects_2024}. The reduction is more noticeable in oxide-based SEs like LLZO, \ce{Li3PO4}, and \ce{Li2O}, where the reduction can reach up to $ \sim 1.3$ eV for LLZO and $ \sim 3$ eV for \ce{Li3PO4}, and \ce{Li2O}. Sulfide-based SEs, such as \ce{Li2S} and \ce{Li3PS4}, exhibit smaller bandgap reduction at grain boundaries (around 1.0 eV or less), whereas halide-based SEs display intermediate changes. These bandgap changes at grain boundaries arise from new localized electronic states due to broken bonds and extended defects. If these states are near the valence band maximum or conduction band minimum, they can significantly enhance electronic leakage.

\begin{figure}[htbp]
    \centering
    \includegraphics[width=1.0\textwidth]{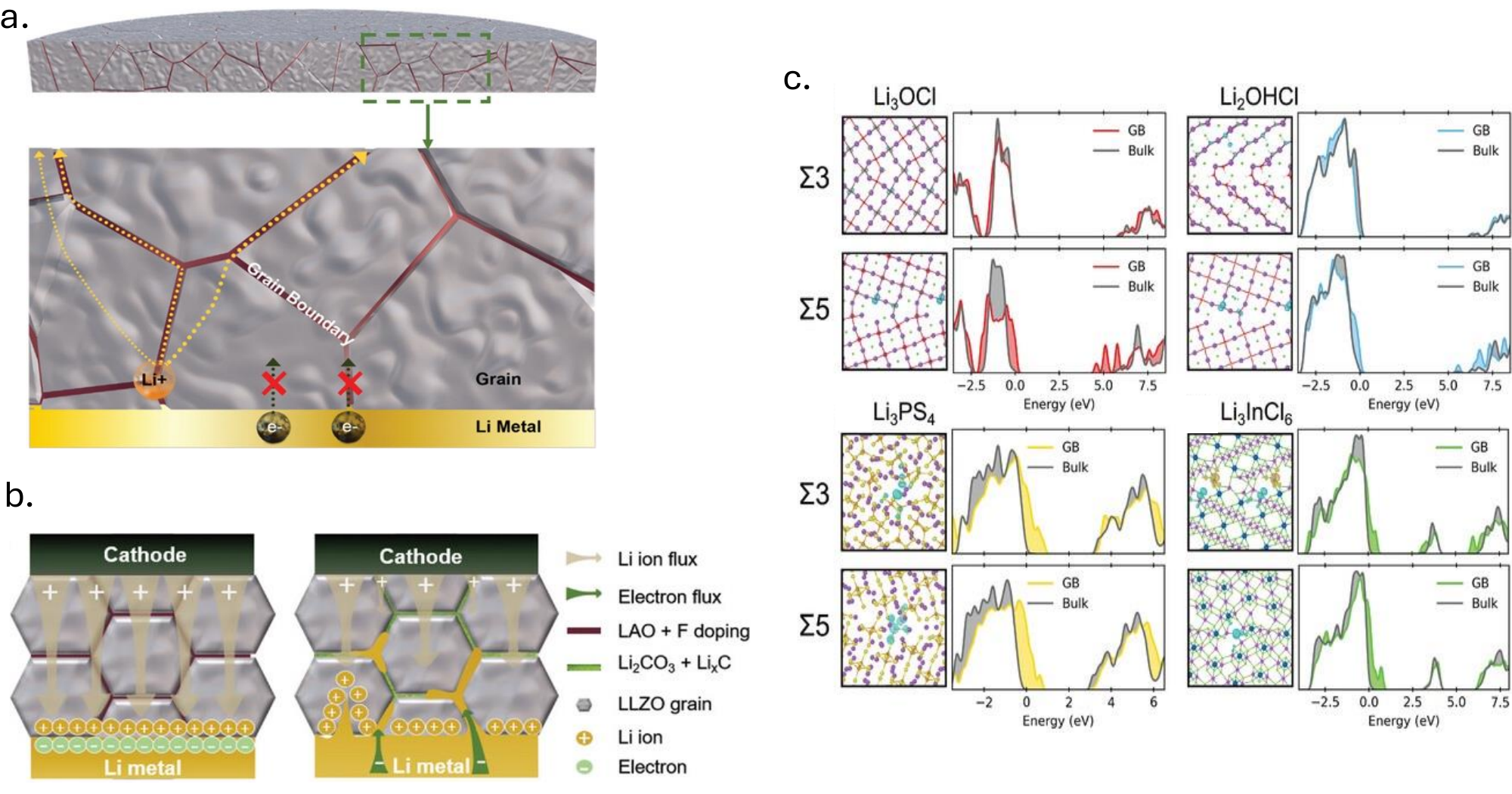}
    \caption{Illustration of mass transport mechanisms in LAO-LLZOF and LLZO.  
(a) The structure of LAO-LLZOF shows multiple pathways for Li-ion movement, while the biphasic grains block electron conduction and  
(b) Schematic of how these materials behave in all-solid-state Li-metal batteries (ASLMBs).  
(left) In LAO-LLZOF, the material allows ion conduction but blocks electrons, promoting even Li plating and stripping during cycling.
(right) In contrast, LLZO allows Li dendrites to form and grow, with filaments extending along electron-conductive grain boundaries and even penetrating the grains. Reproduced with permission from Ref.~\cite{biao_inhibiting_2023}. Copyright \textcopyright2022 2023 Wiley-VCH GmbH.(c) Projected density of states and grain boundary structures for Li$_3$OCl, Li$_2$OHCl, Li$_3$PS$_4$, and Li$_3$InCl$_6$ reveal that grain boundaries, even when having minimal impact on ionic conductivity, can still alter the electronic structure, potentially causing unwanted electronic conductivity and promoting Li dendrite growth. Figure reproduced from \citet{quirk_design_2023} under the terms of the Creative Commons Attribution License. }
    \label{fig:nav-3}
\end{figure}

The ionic and electronic conductivity of LLZO can be tuned by using \ce{Li3AlF6} as an additive during sintering. This leads to 
incorporation of LAO and F-doping at grain boundaries, thereby helping to block dendrite growth (\autoref{fig:nav-3}a and b)\cite{biao_inhibiting_2023}. The incorporation of a continuous, highly conductive LAO phase leads to a dense LAO–LLZOF structure, enabling efficient and uniform Li$^+$ transport through three interconnected pathways: along grain boundaries, from bulk to grain boundaries, and across bulk–grain boundary-bulk regions of LLZO. 
 Introducing an additional anti-perovskite layer at the LLZO/electrode interface is another method to suppress dendrite growth\cite{fu2024anti}.

For halides, there is a risk of increased electronic conductivity driven by trap state-mediated polaron hopping, which can compete with ionic transport\cite{quirk_design_2023}. Innovations like coating nanostructured Li$_3$AlF$_6$ with ionic liquid have enabled more uniform Li flux, smoother deposition, and vastly improved cycling stability\cite{doi:10.1021/acsami.8b12579}.

 In case of sulfide-based SEs, \citet{wang2024key} demonstrated that, in LPSCl, grain boundaries mainly influence electronic transport rather than ion transport. For instance, in LGPS, grain boundaries can develop electron-conductive states at the boundaries\cite{hwang_electrochemical_2023}. Introducing engineered solid electrolyte interphase layers has been shown to effectively block dendrite penetration \cite{wang_all-solid-state_2022,gu2023insights}. In fully amorphous systems, erasing grain boundaries enhances stability and also can suppress dendrite growth\cite{zhao_grainboundaryfree_2022}

Anti-perovskites can also suffer from increased electronic conductivity and a higher risk of dendrite growth at grain boundaries, but these drawbacks can be addressed by optimizing grain size, engineering boundary structures, or using hydrated analogs. \cite{dutra2023computational}. \autoref{fig:nav-3}c compares the density of states for antiperovskite, halide, and sulfide SEs, showing that in all cases, the band gap is significantly reduced at grain boundaries compared to the bulk.

\section{Outlook}\label{sec:outlook}
Grain boundaries play a critical role in determining the performance, safety, and reliability of SSBs, particularly under fast-charging and discharging conditions. A deeper understanding and precise control of grain boundary behavior are essential for realizing the full potential of ceramic SEs.
Despite recent advances, several open questions must be addressed to enable grain boundary-informed design of high-performance SSBs. First, the mechanism of Li dendrite growth from soft to hard short circuit involving grain boundaries remains under debate. It is likely that multiple mechanisms operate depending on the SE chemistry, microstructure, and operating conditions. For example, even within the same SE, dendrite behavior may differ due to processing history or local defect environments. Nevertheless, grain boundaries are a key factor in all observed mechanisms of failure of SSBs.
Second, our understanding of space charge layers near grain boundaries is incomplete. In particular, whether Li enrichment or depletion dominates under different conditions is still unclear. There is a need to bridge atomic-scale insights, such as defect formation and migration at specific grain boundaries, with macroscopic SSB behavior. This requires incorporating realistic factors such as disorder, compositional variations, and electronic leakage into grain boundary models. Advances in space charge layer modeling, particularly those that can isolate grain boundary-specific contributions to ionic conductivity, beyond simplified equivalent circuits and brick layer models,  are needed. Complementary spatially resolved experimental techniques can help reveal heterogeneity within and across grains.
Third, new processing techniques are needed to directly engineer grain boundaries rather than focusing only on electrode/electrolyte interfaces. This is especially critical given emerging evidence that dendrite formation may initiate within the bulk SE, driven by grain boundary-related electronic leakage.
In summary, tackling these challenges will require a tightly integrated approach combining atomistic modeling, mesoscale simulations, advanced synthesis, and spatially resolved characterization. Such efforts will pave the way for grain boundary-optimized SEs for enhanced performance and safety in SSBs.

\begin{acknowledgement}
We acknowledge the Texas Tech University Mechanical Engineering Department startup grant for support of this research. C. W. D. acknowledges support from the Honors College at Texas Tech University.
Z. A. thanks Dr. Jeff Sakamoto and Dr. William Chueh for helpful discussions on grain boundaries and space charge layers in SSBs.

\end{acknowledgement}

\bibliography{zotero,refs}
\end{document}